\documentclass[aps,showpacs,twocolumn,prl,superscriptaddress]{revtex4-2}
\usepackage[english]{babel}
\usepackage{amsmath, amssymb, amsfonts}
\allowdisplaybreaks[4]
\usepackage{amssymb,xcolor}
\usepackage{mathbbol}
\usepackage[colorlinks,linkcolor=blue, urlcolor=blue, citecolor=blue]{hyperref} 
\usepackage{graphicx} 
\usepackage{lipsum} 

\begin{document}
\title{A Necessary and Sufficient Entanglement Criterion of $N$-qubit System Based on Correlation Tensor}
\author{Feng-Lin Wu}
\affiliation{Institute of Modern Physics, Northwest University, Xi'an 710127, China}
\affiliation{Shaanxi Key Laboratory for Theoretical Physics Frontiers, Xi'an 710127, China}
\author{Si-Yuan Liu}
\email{lsy5227@163.com}
\affiliation{Institute of Modern Physics, Northwest University, Xi'an 710127, China}
\affiliation{Shaanxi Key Laboratory for Theoretical Physics Frontiers, Xi'an 710127, China}
\author{Wen-Li Yang}
\affiliation{Institute of Modern Physics, Northwest University, Xi'an 710127, China}
\affiliation{Shaanxi Key Laboratory for Theoretical Physics Frontiers, Xi'an 710127, China}
\author{Shao-Ming Fei}
\affiliation{School of Mathematical Sciences, Capital Normal University, 100048 Beijing, China}
\author{Heng Fan}
\affiliation{Institute of Modern Physics, Northwest University, Xi'an 710127, China}
\affiliation{Shaanxi Key Laboratory for Theoretical Physics Frontiers, Xi'an 710127, China}
\affiliation{Institute of Physics, Chinese Academy of Sciences, Beijing 100190, China}
\affiliation{Beijing Academy of Quantum Information Sciences, Beijing 100193, China }
\affiliation{CAS Center for Excellence in Topological Quantum Computation, UCAS, Beijing 100190, China}
\affiliation{Songshan Lake Materials Laboratory, Dongguan, Guangdong 523808, China}

\date{\today }
\begin{abstract}
Great advances have been achieved in studying characteristics of entanglement for fundamentals of quantum mechanics and quantum information processing.
However, even for $N$-qubit systems, the problem of entanglement criterion has not been well solved.
In this Letter, using the method of state decomposition and high order singular value decomposition (HOSVD), we propose a necessary and sufficient entanglement criterion for general $N$-qubit systems. As an example, we apply our method to study the multi-qubit W state with white noise. We not only obtain the separability critical point,
which is tight and thus better than known results, but also the separable pure state decomposition.
More examples are presented to show that our criterion is accurate, which is tighter than the well-known positive partial transpose criterion.
For two-qubit case, we provide an entanglement measurer which gives similar results with concurrence up to a factor. Our results pave the way to solve the entanglement-separability criterion for more general cases.

\end{abstract}
\pacs{03,65,Ud; 03,67,-a; 03,65,-w}
\maketitle

\textit{Introduction} - As the key quantum resource, entanglement plays an essential role in quantum computation and information processing, for example, such as
one-way quantum computation \cite{PhysRevLett.70.1895, PhysRevLett.119.200501}.
Remarkable progresses have been made in studying various properties of quantum entanglement.
On the other hand, there still exist many important open questions such as the entanglement-separability criterion, particularly for multipartite and high dimensional mixed states.

The study of entanglement has a long history. The concept of entanglement was proposed by Einstein {\it et al.} \cite{einstein1935can} to challenge the completeness of quantum mechanics, while Bell presented the Bell inequality to detect the phenomena of non-locality \cite{bell1964einstein}. Werner proposed that any $N$-partite quantum state $\rho$ is entangled if it can not be written as an ensemble of separable states \cite{werner1989quantum},
\begin{equation} \label{DefEnt}
\rho = \sum_{m} p_m \rho^{(m)}_1 \otimes \rho^{(m)}_2 \otimes \cdots \otimes \rho^{(m)}_N,
\end{equation}
where $\rho^{(m)}_i$ are local density matrices associated with the $i$th partite.
In the past years, a series of entanglement criteria have been proposed.
Peres proposed the positive partial transpose (PPT) criterion, which is proved to be a necessary and sufficient entanglement criterion for two-qubit and qubit-qutrit states \cite{peres1996separability,HORODECKI19961}.
Based on this criterion, the entanglement negativity \cite{PhysRevA.58.883, PhysRevA.65.032314} is shown to be an entanglement monotone. However, the PPT criterion becomes ineffective in identifying entanglement for states other than two-qubit or qubit-qutrit ones. Besides, for two-qubit systems, the concurrence is a well-accepted entanglement measurer
which is computable without difficult optimization approaches \cite{PhysRevLett.78.5022, PhysRevLett.80.2245}. Nevertheless, even for three-qubit states, the situation becomes much more complicated \cite{PhysRevLett.85.1560}.

Entanglement witness is also an important approach in identifying entanglement, see for example \cite{PhysRevLett.92.087902, PhysRevLett.121.200503, PhysRevLett.124.200502, PhysRevLett.127.220501, PhysRevLett.126.140503}. One entanglement witness is the standard entanglement witness proposed in Ref. \cite{sperling2009necessary}, which was further improved in subsequent works \cite{PhysRevLett.118.110502, PhysRevA.97.032343}. But it still does not work for general mixed states. Additionally, various measurers of multipartite entanglement are investigated \cite{PhysRevLett.114.160501, PhysRevLett.117.060501, PhysRevLett.127.140501, PhysRevLett.127.010401}, with some measurers focusing on experimental implementations \cite{guhne2007estimating, PhysRevLett.117.170403}. The genuine multipartite entanglement concurrence was proposed in Refs. \cite{ma2011measure, PhysRevA.92.062338}. And methods of geometry and informatics \cite{zhao2016new, li2013quantifying, xie2021triangle} are also introduced to study entanglement.

In this Letter, we investigate the multipartite entanglement criterion of general $N$-qubit mixed states, which is based on quantum state decomposition and HOSVD. We use correlation tensor to decompose quantum states, and identify whether they can be written as ensemble of separable states or not. We present a necessary and sufficient criterion of separability.

\textit{The Criterion} - Let us first consider a general single qubit state, $\rho=\frac{1}{2}(\mathcal I + \boldsymbol t \cdot \boldsymbol\sigma)$, where $\boldsymbol t$ is a real vector in Bloch sphere, $\mathcal I$ is the identity and $\boldsymbol\sigma$ is the vector given by the Pauli operators. It can be rewritten as a mixture of pure state and completely mixed state, $\rho=(1-|\boldsymbol t|)\frac{1}{2}\mathcal I + |\boldsymbol t|\frac{1}{2}(\mathcal I + \frac{\boldsymbol t}{|\boldsymbol t|}\cdot \boldsymbol\sigma)$. $|\boldsymbol t|$ is the minimal probability related to the pure state. Similarly, the density matrix of an $N$-qubit state can be written as $\rho=\frac{1}{2^N}\sum t_{i_1 i_2 \cdots i_N}\sigma_{i_1}\otimes \sigma_{i_2} \otimes \cdots \otimes \sigma_{i_N}$. Our method
is to analyze the structure of the elements $t_{i_1 i_2 \cdots i_N}$.

For $2$-qubit systems, the elements $t_{0i_2}$ and $t_{i_10}$ are not related the global correlations and do not influence entanglement strength.
We then define the correlation matrix $\mathcal T$ as the absolute values of elements $t_{i_1i_2}$ related to the terms $\frac{1}{4}(\mathcal{I}\pm \sigma_{i_1}\otimes \sigma_{i_2})$, where $i_1,i_2=1,2,3$, see supplementary material Sec. \uppercase\expandafter{\romannumeral1} for details of the correlation matrices $\mathcal R$ and $\mathcal T$. If $\sum_{i_1,i_2=1}^3 |t_{i_1i_2}| \leq 1$, the remained probability of the
density matrix can be allocated to $\frac{1}{4} \mathcal I$. If the minimal sum of $|t_{i_1i_2}|$ was found out,
which is defined up to all possible local unitary transformations, we can easily judge whether a state is entangled or not. Besides, the $\sigma_{i_1}\otimes \sigma_{i_2}$ and the corresponding $|t_{i_1i_2}|$ can be seen as local hidden variables (or states) and their probability. The minimal sum of $|t_{i_1i_2}|$, $S=\min \sum_{U_1\otimes U_2} |t_{i_1i_2}|$, can be obtained by singular value decomposition (SVD) of the correlation matrix $\mathcal T$. And we can prove that $\max(S-1,0)$ is an entanglement measurer of 2-qubit states, see the supplementary material Sec. \uppercase\expandafter{\romannumeral2}.

For general multi-qubit states the minimal probability sum of local hidden variables can be found out based on considerations that any separable states must meet the local hidden variable theory. However, different from the 2-qubit case, the non-global correlations can also influence the entanglement. Considering the local hidden variable theory and the global influence of non-global correlation, we need to preprocess the correlation tensor.

\textit{The Preprocessing of Correlation Tensor} - For the $N$-qubit case, by using Pauli operators, Eq.(\ref{DefEnt}) can be rewritten as
\begin{equation}
  \begin{aligned}
    \rho &=\sum_m [p_m \bigotimes_{n=1}^N \frac{1}{2}(\sum_{i_n=0}^3 t^{(m)}_{i_n}\sigma^{(m)}_{i_n})] \\
         &=\sum_{m} p_m (\frac{1}{2} \prod_{k=i_1}^{i_N} t_k^{(m)} \times \bigotimes_{k=i_1}^{i_N} \sigma_k^{(m)}).
  \end{aligned}
\end{equation}
But in many cases, the global elements of correlation tensor may vanish in the sense that $\sum_m p_m \prod_{k=i_1}^{i_N} t^{(m)}_{k} =0, \quad(i_n\neq 0)$, while the non-global elements remain. To get a strict criterion, we should rebuild the vanishing global elements.

In the state rebuilding, we should obey the law of local hidden variable theory. This classic correlation theory requires that all the components of correlation tensor, especially the non-global elements, cannot be used twice. We have some tables to assist this decomposition-recomposition processing. Taking the 3-qubit $\sigma_3\otimes\sigma_3\otimes\sigma_3$ direction as an example, we have the Tab.\ref{tab1},
where $(j_1 j_2 j_3) = 8|j_1 j_2 j_3\rangle\langle j_1 j_2 j_3|- \mathcal I$ and $|j_1\rangle,|j_2\rangle,|j_3\rangle$ are pure qubit states. It shows the correlation strengths of different directions according to states. The subtracted $\mathcal I$ is to simplify the calculation next. For composing $\bigotimes_{k=i_1}^{i_3} \sigma_k$ merely, there is Tab.\ref{tab2}. And we can expand these tables with $\{|+\rangle,|-\rangle\}$ to $\sigma_1$, $\{|\tilde{+}\rangle,|\tilde{-}\rangle\}$ to $\sigma_2$ and $\{|0\rangle,|1\rangle\}$ to $\sigma_3$, where $|\pm\rangle=\frac{1}{\sqrt{2}}(|0\rangle\pm|1\rangle)$ and $|\tilde{\pm}\rangle=\frac{1}{\sqrt{2}}(|0\rangle\pm \text{i}|1\rangle)$. This processing does not concern with the index $m$. We present an example of a 3-qubit state in supplementary material Sec. \uppercase\expandafter{\romannumeral3}, and show that some elements have first priority in this rebuilding.

\begin{table*}
\centering
\caption{\label{tab1}The decomposition of $(j_1 j_2 j_3)$, where $|j_1\rangle,|j_2\rangle,|j_3\rangle=|0\rangle,|1\rangle$ in this example. Note that $(j_1j_2j_3)= \sum_{i_1,i_2,i_3=0,3} (t_{i_1i_2i_3}\bigotimes_{k=i_1}^{i_3} \sigma_k)$.}
\begin{ruledtabular}
  \begin{tabular}{ccccccccc}
    \hline
    $t_{i_1 i_2 i_3}$ & (000) & (001) & (010) & (011) & (100) & (101) & (110) & (111) \\
    \hline
    $\sigma_3 \otimes \mathcal I \otimes \mathcal I$ & 1 & 1 & 1 & 1 &-1 &-1 &-1 &-1 \\
    $\mathcal I \otimes \sigma_3 \otimes \mathcal I$ & 1 & 1 &-1 &-1 & 1 & 1 &-1 &-1 \\
    $\mathcal I \otimes \mathcal I \otimes \sigma_3$ & 1 &-1 & 1 &-1 & 1 &-1 & 1 &-1 \\
    $\sigma_3 \otimes \sigma_3 \otimes \mathcal I$  & 1 & 1 &-1 &-1 &-1 &-1 & 1 & 1 \\
    $\sigma_3 \otimes \mathcal I \otimes \sigma_3$  & 1 &-1 & 1 &-1 &-1 & 1 &-1 & 1 \\
    $\mathcal I \otimes \sigma_3 \otimes \sigma_3$  & 1 &-1 &-1 & 1 & 1 &-1 &-1 & 1 \\
    $\sigma_3 \otimes \sigma_3 \otimes \sigma_3$   & 1 &-1 &-1 & 1 &-1 & 1 & 1 &-1 \\
    \hline
  \end{tabular}
\end{ruledtabular}
\end{table*}

\begin{table*}
  \centering
  \caption{\label{tab2}The composition of $\bigotimes_{k=i_1}^{i_3} \sigma_k$, where $i_1,i_2,i_3=0,3$ in this example. If the sign of $t_{i_1 i_2 i_3}$ is positive, the composition of $\bigotimes_{k=i_1}^{i_3} \sigma_k$ should be the first column, negative value according to the second column.}
  \begin{ruledtabular}
  \begin{tabular}{ccc}
    \hline
    sign of $t_{i_1 i_2 i_3}$ & + & - \\
    \hline
    $\sigma_3 \otimes \mathcal I \otimes \mathcal I$ & $\frac{t_{300}}{4}[(000)+(001)+(010)+(011)]$ & $\frac{t_{300}}{4}[(100)+(101)+(110)+(111)]$ \\
    $\mathcal I \otimes \sigma_3 \otimes \mathcal I$ & $\frac{t_{030}}{4}[(000)+(001)+(100)+(101)]$ & $\frac{t_{030}}{4}[(010)+(011)+(110)+(111)]$ \\
    $\mathcal I \otimes \mathcal I \otimes \sigma_3$ & $\frac{t_{003}}{4}[(000)+(010)+(100)+(110)]$ & $\frac{t_{003}}{4}[(001)+(011)+(101)+(111)]$ \\
    $\sigma_3 \otimes \sigma_3 \otimes \mathcal I$  & $\frac{t_{330}}{4}[(000)+(001)+(110)+(111)]$ & $\frac{t_{330}}{4}[(010)+(011)+(100)+(101)]$ \\
    $\sigma_3 \otimes \mathcal I \otimes \sigma_3$  & $\frac{t_{303}}{4}[(000)+(010)+(101)+(111)]$ & $\frac{t_{303}}{4}[(001)+(011)+(100)+(110)]$ \\
    $\mathcal I \otimes \sigma_3 \otimes \sigma_3$  & $\frac{t_{033}}{4}[(000)+(100)+(011)+(111)]$ & $\frac{t_{033}}{4}[(001)+(101)+(010)+(110)]$ \\
    $\sigma_3 \otimes \sigma_3 \otimes \sigma_3$   & $\frac{t_{333}}{4}[(000)+(011)+(101)+(110)]$ & $\frac{t_{333}}{4}[(100)+(010)+(001)+(111)]$ \\
    \hline
  \end{tabular}
  \end{ruledtabular}
\end{table*}

Then, we can find out the hidden $t_{i_1i_2i_3}$. These hidden $t_{i_1i_2i_3}$ do not interact with other elements of $\mathcal T$ directly, otherwise there will be some non-physical results. We should make the hidden elements composite an additional tensor $\mathcal T_{\text{add}}$ and use it below.
Define that the element we calculate as $\hat t_{i_1i_2i_3}$, there will be $|t_{i_1i_2i_3}|\leq\hat t_{i_1i_2i_3}$, and $t_{\text{add}.i_1i_2i_3}=\hat t_{i_1i_2i_3}-|t_{i_1i_2i_3}|$.

In a finite $N$-qubit system, the elements of the tensor $\mathcal R$ is finite. Although the intersection set of elements has the first priority in recomposition, we can still use exhaustive method to get the optimal solution, which has the minimal sum of singular values of $\mathcal T_{\text{add}}$. Singular values of tensor will be explained in next section.

\textit{The High Order Singular Value Decomposition (HOSVD)} - With the preprocessing of correlation tensor, we transform the non-global correlations to a kind of equivalent global correlations. To find the minimal probability sum of local hidden variables (or states), we should use some methods of correlation tensor and HOSVD \cite{zhang2013matrix}.

In tensor analysis theory, we can map $\text{SU}(2)^{\otimes N}$ to $\text{SO}(3)^{\otimes N}$ by using tensor $n$-mode product. For qubit case, one has
\begin{footnotesize}
\begin{equation}\label{map1}
  \begin{aligned}
    U_x=\left(
          \begin{array}{cc}
            \cos \frac{\theta}{2} & \text i \sin \frac{\theta}{2} \\
            -\text i \sin \frac{\theta}{2} &  \cos \frac{\theta}{2} \\
          \end{array}
        \right)&\mapsto
    O_x=\left(
          \begin{array}{ccc}
            1 &   &   \\
              & \cos \theta & -\sin \theta \\
              & \sin \theta &  \cos \theta \\
          \end{array}
        \right) \\
    U_y=\left(
          \begin{array}{cc}
            \cos \frac{\theta}{2} & -\sin \frac{\theta}{2} \\
            \sin \frac{\theta}{2} &  \cos \frac{\theta}{2} \\
          \end{array}
        \right)&\mapsto
    O_y=\left(
          \begin{array}{ccc}
            \cos \theta &   & -\sin \theta \\
              & 1 & \\
            \sin \theta &   &  \cos \theta \\
          \end{array}
        \right) \\
    U_z=\left(
          \begin{array}{cc}
            1 &   \\
              &  e^{\text i \theta} \\
          \end{array}
        \right)&\mapsto
    O_z=\left(
          \begin{array}{ccc}
            \cos \theta & -\sin \theta &   \\
            \sin \theta &  \cos \theta &   \\
              &   &  1 \\
          \end{array}
        \right).
  \end{aligned}
\end{equation}
\end{footnotesize}
Under this mapping, we can write a local unitary transformation as,
\begin{small}
\begin{equation}\label{map2}
\begin{aligned}
    &\mathcal I^{\otimes(n-1)} \otimes U^{(n)} \otimes \mathcal I^{\otimes(N-n)} \rho \mathcal I^{\otimes(n-1)} \otimes U^{\dagger{(n)}} \otimes \mathcal I^{\otimes(N-n)} \\
    &\mapsto  \mathcal T \times_n O^{(n)},
\end{aligned}
\end{equation}
\end{small}
where
\begin{equation}
  \begin{aligned}
    \mathcal T \times_n O &= (t_{i_1 \cdots i_N})_{d_1 \cdots d_N} \times_n (o_{m i_n})_{d_m d_n} \\
    &=\sum_{i_n} t_{i_1 \cdots i_n \cdots i_N}o_{m i_n} \\
    &=\mathcal (t'_{i_1 \cdots i_{n-1} m i_{n+1} \cdots i_N})_{d_1 \cdots d_{n-1} d_m d_{n+1} \cdots d_N},
  \end{aligned}
\end{equation}
with the index $n$ denoting the $n$-th partite.
In the case of $N$-qubit system, $d_1,d_2, \cdots, d_N =3$, $O\in \mathbb R^{3\times 3}$.

Denote $\langle \mathcal A, \mathcal B \rangle=\sum_{i_1,i_2,\cdots,i_N} a_{i_1 i_2 \cdots i_N}b^*_{i_1 i_2 \cdots i_N}$ the inner product of tensors $\mathcal A$ and $\mathcal B$.
For the $N$-qubit case, what we need is that
\begin{small}
\begin{equation}\label{independent}
    \langle \mathcal T'_{i_1 \cdots i_{k-1} k_1 i_{k+1} \cdots i_N},\mathcal T'_{i_1 \cdots i_{k-1} k_2 i_{k+1} \cdots i_N} \rangle
    =0,~~ \forall k_1 \neq k_2,
\end{equation}
\end{small}
which can be solved perfectly by HOSVD \cite{zhang2013matrix}.

As for HOSVD, an $I_1 \times I_2 \times \cdots \times I_N$ real tensor $\mathcal X$ can be decomposed as
\begin{equation}
\mathcal X= \mathcal G \times_1 P^{(1)} \times_2 P^{(2)} \cdots \times_N P^{(N)},
\end{equation}
where $P^{(n)}=[\boldsymbol p_1^{(n)},\cdots,\boldsymbol p_{J_n}^{(n)}]$ is a semi-positive orthogonal matrix, with ${P^{(n)}}^T P^{(n)}=\mathcal I_{J_n}$, $I$ and $J$ are the dimensions of different directions, $P$ is a matrix of row $I$ and column $J$, $\boldsymbol p^{(n)}_{j_n}$ are $I_n$ dimensional vector. The core tensor $\mathcal G$ is a $J_1\times J_2\times \cdots \times J_N$ tensor, whose subtensor $\mathcal G_{j_n=\alpha}$ is the tensor $\mathcal X$ which makes the index $j_n=\alpha$ be fixed. The core tensor has two features, the one is all-orthogonality that any $\mathcal G_{j_n=\alpha}$ and $\mathcal G_{j_n=\beta}$ satisfy
\begin{equation}
  \langle \mathcal G_{j_n=\alpha},\mathcal G_{j_n=\beta}\rangle =0, \,\,\, \forall\alpha \neq \beta,
\end{equation}
and the other one is sorted by
\begin{equation}
  \|\mathcal G_{j_n=1} \|_{\text F} \geq \|\mathcal G_{j_n=2} \|_{\text F} \geq \cdots \geq \|\mathcal G_{j_n=N} \|_{\text F},
\end{equation}
where $\| \mathcal A \|_{\text F}$ is the Forbenius norm of tensor $\mathcal A$.

Decomposing the correlation tensor $\mathcal T$ by using HOSVD, we obtain
\begin{equation}
\mathcal T = \mathcal T' \times_1 P^{(1)} \times_2 P^{(2)} \cdots \times_N P^{(N)},
\end{equation}
where $P^{(1)}, P^{(2)}, \cdots, P^{(N)}$ are O(3) operators. But due to Eq.(\ref{map1}), here we should concern SO(3) operators. Fortunately, as SO(3) operators belong to O(3) operators, their differences just influence the signs of $t'_{i_1 i_2 \cdots i_N}$, and we need to use $|t'_{i_1 i_2 \cdots i_N}|$. In fact, it is not a trouble.

Now we have a tensor $\mathcal T'$ with orthogonal slices. This means that these
slices are independent with each other. We can regard each slice as a quantum state,
and operate every slice independently.
In the general case, every slice is orthogonal to each other. The simplest superposition is to superpose every slice (in a specific mode which is feasible in physics) as an independent quantum state. It means that to calculate the minimal value of $\sum_k |t'_k|$ is to calculate the summation of the singular values of all the slices. To ensure this decomposition is a physics processing, any singular value of each slice must be less than or equal to $1$ in feasible segmentation mode.

Although the subtensors (slices) of the tensor $\mathcal T'$ are orthogonal, the slices of subtensors are not always orthogonal. We should repeat above HOSVD and superposition to get $\mathcal T'^{(2)}$, $\mathcal T'^{(3)}$, $\cdots$. Every time we repeat this processing, the subtensors will reduce 1 order. The process repeats till the subtensors become $3$-dimensional square matrices with order 2. And then, there will be $3^{N-2}$ $3$-dimensional square matrices with minimal value of $\sum_k |t'^{(N-2)}_k|$ (superscripts like $(N-2)$ are omitted unless otherwise specified, $\mathcal T'$ and $t'$ represent the result of iteration in the last step). Calculating their
singular values $\{s^1,s^2,s^3\}_{I_1I_2\cdots i_k\cdots i_l\cdots I_N}$ and finding out the decomposition mod, we have
\begin{equation}
  S_{\text{min}}=\min_{i_k,i_l}\sum_{r,k,l} s^r_{I_1I_2\cdots i_k\cdots i_l\cdots I_N},
\end{equation}
where $r$ is the number of singular values with each slice, $I$'s are the fixed indices.
In this case, we can rearrange the singular values as a new tensor $\mathcal S$.
If $s_{i_1i_2\cdots i_N} \neq 0$,
\begin{equation}\label{LimCon}
  s_{i_1i_2\cdots i_{n-1} j_n i_{n+1} \cdots i_N}=\delta_{i_nj_n} s_{i_1i_2\cdots i_N},
\end{equation}
where $j_n$ could be any partite. This form is in consistent with Eq.(\ref{independent}). We show an example of 3-qubit $\mathcal S$ which is rearranged in supplementary material Sec. \uppercase\expandafter{\romannumeral4}.

By using the same method to $\mathcal T_{\text{add}}$, we will get $S_{\text{add.min}}$. Now we define
\begin{equation}
  S=\sum s+\sum s_{\text{add}}.
\end{equation}
We have the following separability criterion,
\begin{equation}
  \left\{
  \begin{aligned}
    &S   >  1,   \,\,\,  \text{state $\rho$ is entangled}, \\
    &S \leq 1,   \,\,\,  \text{state $\rho$ is separable},
  \end{aligned}
  \right.
\end{equation}
where $S$ represents the minimal probability sum of separable states except for the completely mixed state.

It is obvious that any quantum state satisfying $S \leq 1$ is separable.
Besides, since the Bloch vector of any qubit state has module less than or equal to 1, its module can be regarded as a probability without completely mixed state. For $N$-qubit state this probability corresponds to a local hidden variable (or state), and our method can find out the minimal sum of probability by using singular value decomposition. This minimal value means that the criterion is a necessary and sufficient separability condition for general $N$-qubit systems. Note that for separable states, the sum of probability can not be greater than 1. The reason is that if a quantum state has $S>1$, one can not find a set of local hidden variables (or states) to construct this state.

\textit{Examples}\label{example} - We take some cases to demonstrate how the criterion works. The multi-qubit W state is shown below. The Greenberger-Horne-Zeilinger (GHZ) state and entanglement sudden death are shown in supplementary material Sec.\uppercase\expandafter{\romannumeral5}, Sec.\uppercase\expandafter{\romannumeral6} and Sec.\uppercase\expandafter{\romannumeral8}. These results illustrate that our method is operational and effective with physical manifestation.

The 3-qubit W state has the form, $|\psi\rangle=\frac{1}{\sqrt{3}}(|001\rangle+|010\rangle+|100\rangle)$. By straightforward calculation we have $S=\frac{19}{3}$, see the details in supplementary material Sec.\uppercase\expandafter{\romannumeral7}.

\begin{figure}[htbp]
\centering
\includegraphics[width=0.45\textwidth]{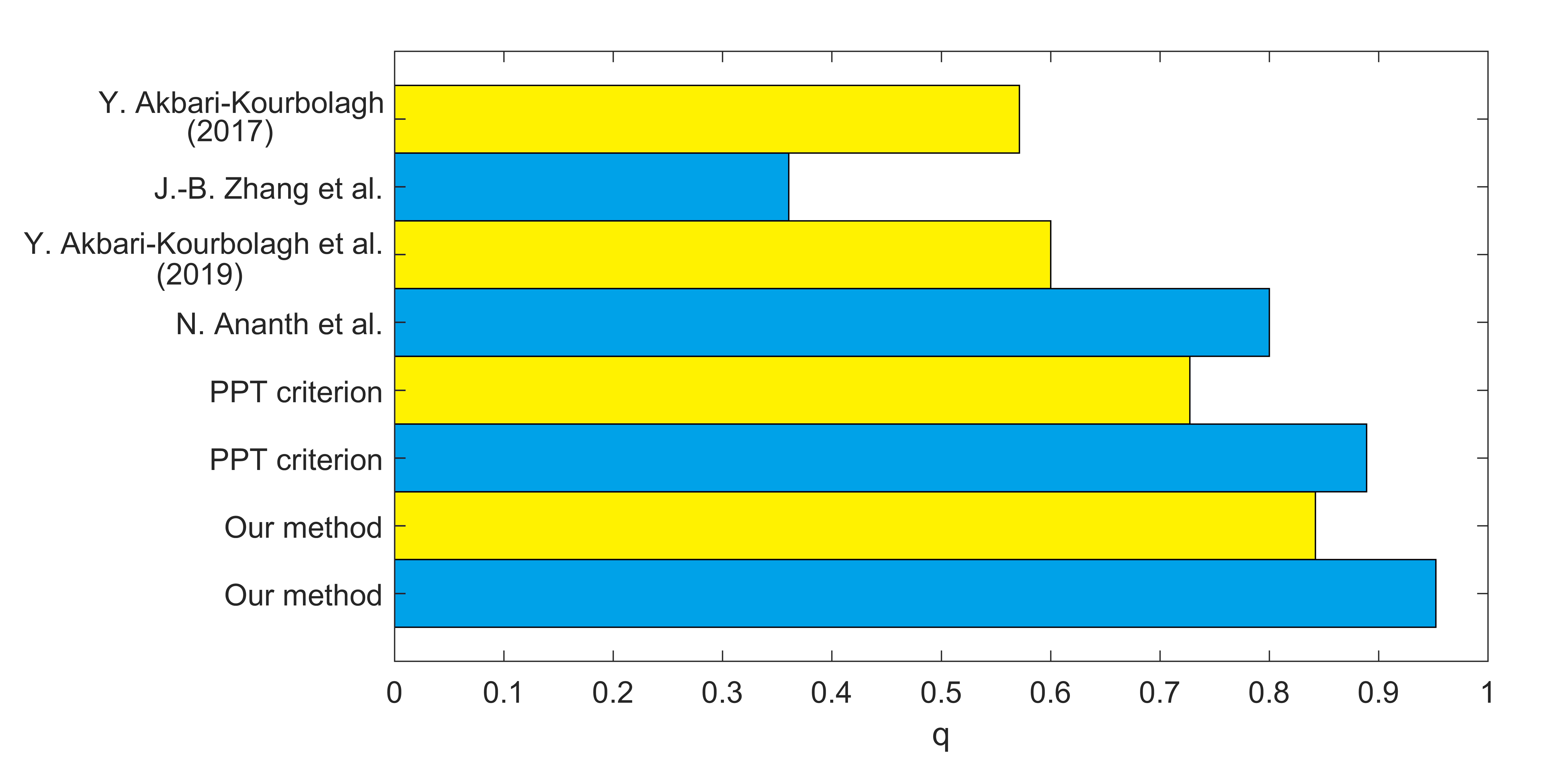}
\caption{Results from different criteria for 3-qubit (yellow bars) and 4-qubit (blue bars) W state mixed white noise with strength $q$ of white noise. The yellow bars represent the regions of $q$ such that the state is entangled, with the results given by Y. Akbari-Kourbolagh ($q<\frac{4}{7}\approx0.5714$ (2017), $q<0.6001$ (2019)), PPT criterion ($q<\frac{8}{11}\approx0.7273$) and our method ($q<\frac{16}{19}\approx0.8421$), respectively.
The blue bars represent the entangled regions of $q$ given by J.-B. Zhang ($q\leq0.3605$), N. Ananth {\it et al.} ($q<\frac{4}{5}=0.8000$), PPT criterion ($q<\frac{8}{9}\approx0.8889$) and our method ($q<\frac{20}{21}\approx0.9524$), respectively.}
\label{34-w}
\end{figure}

Now consider the mixture of the 3-qubit W state and white noise with strength $q$, $\hat{\rho}=(1-q)\rho+q\frac{\mathcal I}{8}$,
where $\rho=|\psi\rangle\langle\psi|$. Then our criterion gives that $\hat{\rho}$ is entangled when $q<\frac{16}{19}\approx0.8421$, which is more precise than the result $q<\frac{4}{7}\approx0.5714$ given in \cite{akbari2017entanglement} and $q<0.6001$ from a criterion based on quantum Fisher information \cite{akbari2019entanglement}.
Due to the symmetry of $\hat{\rho}$, one may also apply the PPT criterion. Our method is also more precise than the result $q<\frac{8}{11}\approx0.7273$ from PPT
criterion \cite{peres1996separability, HORODECKI19961} (see Fig.\ref{34-w}).


Similarly, we can give the result for the mixture of the 4-qubit W state, $|\psi\rangle=\frac{1}{2}(|0001\rangle+|0010\rangle+|0100\rangle+|1000\rangle)$, and white noise with strength $q$. We get $S=21$ and the state is entangled when $q<\frac{20}{21}\approx0.9524$. This result is more precise than $q\leq0.3605$ give by J.-B. Zhang {\it et al.} \cite{zhang2021multipartite}, $q<\frac{4}{5}=0.8000$ given by N. Ananth {\it et al.} \cite{ananth2015criteria} and $q<\frac{8}{9}\approx0.8889$ from the PPT criterion \cite{peres1996separability, HORODECKI19961} (See Fig.\ref{34-w}).

In addition to the precise separability, our method also gives rise to the detailed separable ensemble when the state is separable. For the case of the mixture of the 3-qubit W state and white noise with strength $q$, when $q=\frac{16}{19}$ we have an ensemble of separable states with $p_1=p_2=p_3=p_4=p_5=p_6=p_7=p_8=\frac{2}{19}$, $p_9=\frac{3}{19}$ and
\begin{footnotesize}
\begin{equation}
  \begin{aligned}
    \rho_1 = &\frac{1}{8}(\mathcal I +\sigma_1\otimes\sigma_1\otimes\sigma_3),~~
    \rho_2 = \frac{1}{8}(\mathcal I +\sigma_1\otimes\sigma_3\otimes\sigma_1), \\
    \rho_3 = &\frac{1}{8}(\mathcal I +\sigma_3\otimes\sigma_1\otimes\sigma_1),~~
    \rho_4 = \frac{1}{8}(\mathcal I +\sigma_2\otimes\sigma_2\otimes\sigma_3), \\
    \rho_5 = &\frac{1}{8}(\mathcal I +\sigma_2\otimes\sigma_3\otimes\sigma_2),~~
    \rho_6 = \frac{1}{8}(\mathcal I +\sigma_3\otimes\sigma_2\otimes\sigma_2), \\
    \rho_7 = &\frac{1}{8}(\mathcal I +\sigma_1\otimes\sigma_1\otimes \mathcal I +\sigma_1\otimes\mathcal I\otimes \sigma_1 +\mathcal I\otimes\sigma_1\otimes \sigma_1), \\
    \rho_8 = &\frac{1}{8}(\mathcal I +\sigma_2\otimes\sigma_2\otimes \mathcal I +\sigma_2\otimes\mathcal I\otimes \sigma_2 +\mathcal I\otimes\sigma_2\otimes \sigma_2), \\
    \rho_9 = &\frac{1}{8}(\mathcal I +\frac{1}{3}\sigma_3\otimes\mathcal I\otimes\mathcal I +\frac{1}{3}\mathcal I\otimes\sigma_3\otimes\mathcal I+\frac{1}{3}\mathcal I\otimes\mathcal I\otimes\sigma_3 \\
             &-\frac{1}{3}\sigma_3\otimes\sigma_3\otimes \mathcal I -\frac{1}{3}\sigma_3\otimes\mathcal I\otimes \sigma_3 -\frac{1}{3}\mathcal I\otimes\sigma_3\otimes \sigma_3 \\
             &-\sigma_3\otimes\sigma_3\otimes\sigma_3).
  \end{aligned}
\end{equation}
\end{footnotesize}
Set $|\pm\rangle=\frac{1}{\sqrt{2}}(|0\rangle\pm|1\rangle$ and $|\tilde{\pm}\rangle=\frac{1}{\sqrt{2}}(|0\rangle\pm \text{i}|1\rangle)$. The above ensemble can be written as a pure state decomposition with $p_1=p_2=\cdots=p_{24}=\frac{1}{38}$, $p_{25}=p_{26}=\cdots=p_{31}=\frac{1}{19}$ and
\begin{equation}
  \begin{aligned}
    &|\psi_{1}\rangle=|\!+\!+0\rangle,\; |\psi_{2}\rangle=|\!+\!-1\rangle,\; |\psi_{3}\rangle=|\!-\!+1\rangle, \\ &|\psi_{4}\rangle=|\!-\!-0\rangle,\; |\psi_{5}\rangle=|\!+\!0+\rangle,\; |\psi_{6}\rangle=|\!+\!1-\rangle, \\ &|\psi_{7}\rangle=|\!-\!0-\rangle,\; |\psi_{8}\rangle=|\!-\!1+\rangle,\; |\psi_{9}\rangle=|0\!+\!+\rangle, \\ &|\psi_{10}\rangle=|0\!-\!-\rangle,\; |\psi_{11}\rangle=|1\!+\!-\rangle,\; |\psi_{12}\rangle=|1\!-\!+\rangle, \\
    &|\psi_{13}\rangle=|\tilde + \tilde + 0\rangle,\; |\psi_{14}\rangle=|\tilde + \tilde - 1\rangle,\; |\psi_{15}\rangle=|\tilde - \tilde + 1\rangle, \\
    &|\psi_{16}\rangle=|\tilde - \tilde - 0\rangle,\; |\psi_{17}\rangle=|\tilde + 0 \tilde +\rangle,\; |\psi_{18}\rangle=|\tilde + 1 \tilde -\rangle, \\
    &|\psi_{19}\rangle=|\tilde - 0 \tilde -\rangle,\; |\psi_{20}\rangle=|\tilde - 1 \tilde +\rangle,\; |\psi_{21}\rangle=|0 \tilde + \tilde +\rangle, \\
    &|\psi_{22}\rangle=|0 \tilde- \tilde-\rangle,\; |\psi_{23}\rangle=|1 \tilde + \tilde -\rangle,\; |\psi_{24}\rangle=|1 \tilde - \tilde +\rangle, \\
    &|\psi_{25}\rangle=|\!+\!++\rangle,\; |\psi_{26}\rangle=|\!-\!--\rangle, \\
    &|\psi_{27}\rangle=|\tilde + \tilde + \tilde +\rangle,\; |\psi_{28}\rangle=|\tilde - \tilde - \tilde -\rangle, \\
    &|\psi_{29}\rangle=|001\rangle,\; |\psi_{30}\rangle=|010\rangle,\; |\psi_{31}\rangle=|100\rangle.
  \end{aligned}
\end{equation}
The separable ensemble for the above 4-qubit case is shown in supplementary material Sec.\uppercase\expandafter{\romannumeral7}.
These results mean that our method can not only detect the entanglement of qubit systems, but also provide an available separable state ensemble.

\textit{Conclusions} - In this paper, we have first time proposed a necessary and sufficient entanglement criterion for general $N$-qubit systems based on high order singular value decomposition. With this method, we can not only judge whether a state is separable or not, but also obtain the optimal decomposition of separable states -- the minimal sum of separable state probability except for the completely mixed state.


We have provided a series of examples to illustrate our method. These examples show that our criterion for general $N$-qubit states is obviously more accurate than previous criterions. Our criterion has explicit physical implications and can explain some phenomenon which is hidden in separable states. Especially, our criterion can be used as an entanglement measurer for 2-qubit systems, from which we can study the entanglement robustness and entanglement sudden death of multi-qubit GHZ states with manifest physical meanings, see supplementary material Sec.\uppercase\expandafter{\romannumeral8}. Our results may highlight further investigations on detection of quantum entanglement for high dimensional systems.


\begin{acknowledgements}
We acknowledge Xiao-Yu Zhang for helpful discussions and Pei Wang for company and encouragement. This work is supported by the National Natural Science Foundation of China (NSFC) under grant Nos. 12075159, 12171044, 12175179, 11934018, Z2121001; Beijing Natural Science Foundation (Grant No. Z190005); Shenzhen Institute for Quantum Science and Engineering, Southern University of Science and Technology (Grant No.SIQSE202001); the Academician Innovation Platform of Hainan Province, and the Peng Huaiwu Center for Fundamental Theory under Grant No.12047502.

\end{acknowledgements}

\bibliography{reference.bib}

\newpage
\setcounter{equation}{0}
\onecolumngrid
\appendix
\renewcommand{\theequation}{S.\arabic{equation}}
\renewcommand{\thefigure}{S.\arabic{figure}}

\begin{center}
\textbf{Supplementary Material for\\~\\A Necessary and Sufficient Entanglement Criterion of $N$-qubit System Based on Correlation Tensor}
\end{center}

\section{\uppercase\expandafter{\romannumeral1} The Correlation Matrix and Tensor}
Any density matrix of a qubit state can be written as $\rho=\frac{1}{2}(t_0\mathcal{I} +t_1\sigma_1 +t_2\sigma_2 +t_3\sigma_3) =\frac{1}{2}(\sum_{i=0}^{3} t_i\sigma_i)$, where $t_0=1$. Thus any density matrix of an $N$-qubit quantum state can be written as
\begin{equation}
  \rho=\frac{1}{2^N}\sum_{i_1,i_2,\cdots,i_N} t_{i_1 i_2 \cdots i_N}\sigma_{i_1}\otimes \sigma_{i_2} \otimes \cdots \otimes \sigma_{i_N},
\end{equation}
where $i_n=0,1,2,3$, $t_{00\cdots0}=1$,
\begin{equation}
\begin{aligned}
  \sigma_0=\mathcal{I}=\left(
             \begin{array}{cc}
               1 &   \\
                 & 1 \\
             \end{array}
           \right),~~~
  \sigma_1=\left(
             \begin{array}{cc}
                 & 1 \\
               1 &   \\
             \end{array}
           \right), \\
  \sigma_2=\left(
             \begin{array}{cc}
                 & -\text i \\
               \text i  &   \\
             \end{array}
           \right),~~~
  \sigma_3=\left(
             \begin{array}{cc}
               1 &   \\
                 & -1 \\
             \end{array}
           \right).
\end{aligned}
\end{equation}

Then we can get an $N$-order $4$-dimensional tensor $\mathcal{R}=(t_{i_1 i_2 \cdots i_N})_{4^{N}}$, $i_n=0,1,2,3$ and an $N$-order $3$-dimensional tensor $\mathcal{T}=(t_{i_1 i_2 \cdots i_N})_{3^{N}}$, $i_n=1,2,3$ related to the correlations among the qubits. We can call $\mathcal{T}$ the correlation tensor. For the 2-qubit case, $\mathcal{T}=(t_{i_1 i_2})_{3\times 3}$ reduces to the correlation matrix. Due to the symmetry and unitarian of the Pauli operators, one has $t_{i_1 i_2 \cdots i_N} =\text{Tr} (\sigma_{i_1} \otimes \sigma_{i_2} \otimes \cdots \otimes \sigma_{i_N} \rho)$.

For $2$-qubit systems, the absolute values of the elements $t_{i_1i_2}$ of the correlation matrix $\mathcal T$ can be regarded as the probability of the state $\frac{1}{4}(\mathcal{I}\pm \sigma_{i_1}\otimes \sigma_{i_2})$. If $\sum |t_{i_1i_2}| \leq 1$, the remained probability can be allocated to $\frac{1}{4} \mathcal I$. Therefore, if we can find some local unitary operations such that $S=\min_{U_1\otimes U_2}\sum |t_{i_1i_2}|$, then $S$ can be used to characterize entanglement. Fortunately, the singular value decomposition can be applied to find the minimal value of $\sum |t_{i_1i_2}|$, which just gives rise to the precise entanglement criterion for $2$-qubit pure states. This singular value matrix, which has only one element in each raw and column, gives the minimal sum of separable state probability except for the completely mixed state. We generalize this method to $N$-qubit system below.

\section{\uppercase\expandafter{\romannumeral2} Proof of the Criterion as an Entanglement Measurer for 2-qubit Systems}
A 2-qubit entanglement measure $E$ should at least satisfy:

M1) $E(\rho)=0$ if and if $\rho$ is separable.

M2) $E(\rho)>0$ if $\rho$ is entangled.

M3) $E(\sum_k p_k \rho_k) \leq \sum_k p_k E(\rho_k)$.

M4) $E([U_1\otimes U_2] \rho [U_1\otimes U_2]^\dagger)=E(\rho)$.

M5) $\sum_k \text{Tr}(V_k\rho V_k^\dagger)E(\frac{V_k\rho V_k^\dagger}{\text{Tr}(V_k\rho V_k^\dagger)}) \leq E(\rho)$, where $\{V_k\}$ are the local operation and classical communication (LOCC) operators.

For the 2-qubit system, the correlation tensor is reduced to be a $3\times 3$ matrix. Let $E_{2\text{-qubit}}=\max (S-1,0)$. It is obvious that the M1) and M2) are satisfied. As the orthogonal operator does not change the singular value of the correlation matrix, with the Eq.3 and Eq.4 in the main text, the condition M4) is also satisfied.

To prove the M3), consider two 2-qubit states $\rho_1$ and $\rho_2$ with $\mathcal S_1$ and $\mathcal S_2$, respectively. We have $\mathcal T_1= O_1 \mathcal S_1 P^\dagger_1$ and $\mathcal T_2= O_2 \mathcal S_2 P^\dagger_2$, where $O$ and $P$ are special orthogonal operators.

Taking $\rho=p_1\rho_1 + p_2\rho_2$ we have
\begin{equation}
  \begin{aligned}
    \mathcal T =& p_1\mathcal T_1 +p_2\mathcal T_2 \\
               =& p_1O_1 \mathcal S_1 P^\dagger_1 + p_2 O_2 \mathcal S_2 P^\dagger_2 \\
               =& O_1 (p_1 \mathcal S_1 + p_2O' \mathcal S_2 P'^\dagger) P^\dagger_1,
  \end{aligned}
\end{equation}
where the closure of orthogonal group has been used in the third equal sign, $O'=O^\dagger_1O_2$ and $P'=P^\dagger_1P_2$ are special orthogonal operators.
Because of the M4), one has
\begin{equation}
  SSV(p_1\mathcal S_1 + p_2O' \mathcal S_2 P'^\dagger) \leq SSV(p_1\mathcal S_1 + p_2\mathcal S_2),
\end{equation}
where $SSV(A)$ stands for the sum of the singular values of the matrix $A$. The less than or equal sign is due to that the squared sum of matrix singular values is equal to the squared sum of all element modules, and all elements are real numbers which are less than or equal to $1$. The equality holds if and only if $O1=O2$ and $P_1=P_2$. And the condition M3) is proved.

For M5), the average of entanglement does not increase under LOCC, and LOCC is equal to all local positive maps. Suppose a local positive map $\Lambda(\rho)=\Lambda_{i_1}\otimes\Lambda_{i_2}(\rho)$.
The local maps in correlation matrix can be regarded as a combination of SO(3) operations (see Eq.3 in the main text) and scaling operations. Generally, we can write the correlation matrix as
\begin{equation}
  \Lambda'(O_1\left(
            \begin{array}{ccc}
              t_{11} & t_{12} & t_{13} \\
              t_{21} & t_{22} & t_{23} \\
              t_{31} & t_{32} & t_{33} \\
            \end{array}
          \right)
  O_2^\dagger) \mapsto
  \left(
    \begin{array}{ccc}
      \eta_1\mu_1t'_{11} & \eta_1\mu_2t'_{12} & \eta_1\mu_3t'_{13} \\
      \eta_2\mu_1t'_{21} & \eta_2\mu_2t'_{22} & \eta_2\mu_3t'_{23} \\
      \eta_3\mu_1t'_{31} & \eta_3\mu_2t'_{32} & \eta_3\mu_3t'_{33} \\
    \end{array}
  \right),
\end{equation}
where $\Lambda'$ is the scaling operation which acts in the $\sigma_k$ direction on the first and second qubits, $\eta_{i_1}$ and $\mu_{i_2}$ are the scaling coefficients. The positive maps $\Lambda(\rho)=\Lambda'((U_1\otimes U_2)\rho (U_1\otimes U_2)^\dagger)$ always map one state to another state, and $|t_{i_1i_2}|$ is always less than or equal to 1 in any quantum state. To ensure that the new matrix is still a quantum state, $|\eta_{i_1}|$ and $|\mu_{i_2}|$ must be less than or equal to 1. And because the squared sum of the matrix singular values is equal to the squared sum of all element modules, the sum of matrix singular values is always non-increasing. Thus M5) is proved.

Below we take the entanglement negativity and concurrence for comparisons.

\subsection{The calculation of negativity of Bell diagonal states}
The negativity \cite{PhysRevA.58.883, PhysRevA.65.032314} has the form,
\begin{equation}\label{neg2}
E_{\text{Negativity}}(\rho)=\text{Tr}(\rho^{T_A}\rho^{T_A\dagger})-1,
\end{equation}
where $\rho^{T_A}$ is the partially transposed matrix of $\rho$ with respect to the first subsystem. We use $E_{\text{Neg-}2}$ to represent negativity for the $2$-qubit case.

Any two-qubit state can be written in the following form, 
\begin{equation}
  \rho=\frac{1}{4}(\mathcal I \otimes \mathcal I +\sum t_{i_10}\sigma_{i_1}\otimes\mathcal I +\sum t_{0i_2} \mathcal I \otimes\sigma_{i_2} +\sum t_{i_1i_2} \sigma_{i_1}\otimes \sigma_{i_2}).
\end{equation}
Under suitable local unitary transformation, the correlation matrix of any
two-qubit state can be written in a diagonal form $\text{diag}\{a,b,c\}$, 
and the state becomes
\begin{equation}\label{BellState}
  \begin{aligned}
    \rho=\frac{1}{4}(\mathcal I\otimes \mathcal I+ a\sigma_1\otimes \sigma_1 +b\sigma_2\otimes \sigma_2 +c\sigma_3\otimes \sigma_3),
  \end{aligned}
\end{equation}
which is the well known Bell-diagonal state. The parameters $a$, $b$ and $c$ of the matrix $\mathcal T$ belong to the tetrahedron with four vertices $v_k=(a_{0_k},b_{0_k},c_{0_k})$, where $v_1=(-1,-1,-1)$, $v_2=(-1,1,1)$, $v_3=(1,-1,1)$ and $v_4=(1,1,-1)$.

Substituting Eq.\ref{BellState} into Eq.\ref{neg2}, we have
\begin{equation}
    E_{\text{Neg-}2}=\frac{|a+b-c-1|}{4}+\frac{|a-b+c-1|}{4}
             +\frac{|-a+b+c-1|}{4}
             +\frac{|-a-b-c-1|}{4} -1.
\end{equation}
From the tetrahedron relation given in Ref. \cite{PhysRevA.54.1838}, there are four cases of entanglement,
\begin{equation}
  E_{\text{Neg-}2}=\left\{
  \begin{aligned}
    -&\frac{a}{2}+\frac{b}{2}+\frac{c}{2}-\frac{1}{2}, \,\,\, a<0,\,b>0,\,c>0, \\
    &\frac{a}{2}-\frac{b}{2}+\frac{c}{2}-\frac{1}{2}, \,\,\, a>0,\,b<0,\,c>0, \\
    &\frac{a}{2}+\frac{b}{2}-\frac{c}{2}-\frac{1}{2}, \,\,\, a>0,\,b>0,\,c<0, \\
    -&\frac{a}{2}-\frac{b}{2}-\frac{c}{2}-\frac{1}{2}, \,\,\, a<0,\,b<0,\,c<0.
  \end{aligned}
  \right.
\end{equation}
Obviously, $E_{\text{Neg-}2}$ can be simplified to be
\begin{equation}
  E_{\text{Neg-}2}=\frac{|a|}{2}+\frac{|b|}{2}+\frac{|c|}{2}-\frac{1}{2}.
\end{equation}
For separable states, the constraint condition $|a|+|b|+|c|\leq 1$ gives that $E_{\text{Neg-}2}=0$. In general we have
\begin{equation}
  E_{\text{Neg-}2}=\max(0, \frac{|a|}{2}+\frac{|b|}{2}+\frac{|c|}{2}-\frac{1}{2}).
\end{equation}

As the sum of the singular values of the correlation matrix,
\begin{equation}
  \mathcal T = \left(
                 \begin{array}{ccc}
                   a & 0 & 0 \\
                   0 & b & 0 \\
                   0 & 0 & c \\
                 \end{array}
               \right),
\end{equation}
is given by
\begin{equation}
  S=|a|+|b|+|c|,
\end{equation}
we have
\begin{equation}
  \begin{aligned}
    E_{2\text{-qubit}}=&\max (S-1,0) \\
                      =&\max (|a|+|b|+|c|-1,0).
  \end{aligned}
\end{equation}
Clearly, $E_{2\text{-qubit}}=2E_{\text{Neg-}2}$, if we normalize the $E_{2\text{-qubit}}$ so that its maximal value is 1. These two measurers are the same for Bell diagonal states.

\subsection{The concurrence and Werner states}
For the Werner state $\rho=(1-q)|\psi^-\rangle\langle\psi^-|+q\frac{\mathcal I}{4}$, where $|\psi^-\rangle=\frac{1}{\sqrt2}(|01\rangle-|10\rangle)$, the concurrence \cite{PhysRevLett.78.5022, PhysRevLett.80.2245} has the form, $E_c=\max(0,\lambda_1-\lambda_2-\lambda_3-\lambda_4)$, where $\lambda_k$ are the eigenvalues of the matrix $R_c=\sqrt{\sqrt{\rho}\tilde{\rho}\sqrt{\rho}}$ with $\tilde{\rho}=(\sigma_y\otimes\sigma_y)\rho^*(\sigma_y\otimes\sigma_y)$, $\lambda_1$ is the largest eigenvalue.

\begin{figure}[htbp]
\centering
\includegraphics[width=0.50\textwidth]{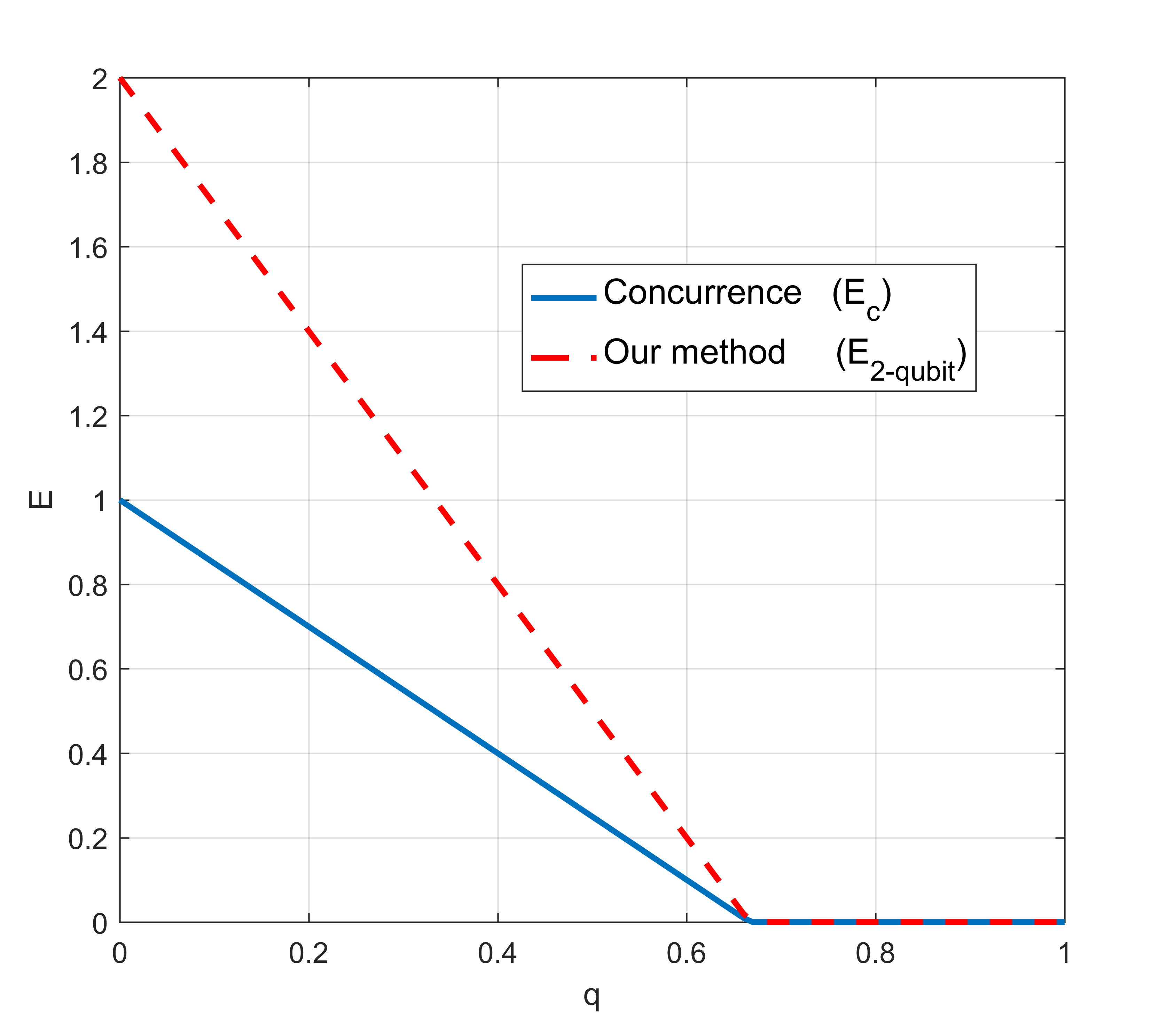}
\caption{$E_{2\text{-qubit}}$ and concurrence for Werner states. They are essentially the same under normalization.}
\label{OursCon}
\end{figure}

The results of our $E_{2\text{-qubit}}$ and concurrence are shown in Fig.\ref{OursCon}. We can find that these two entanglement measurers have similar behaviors.

\section{\uppercase\expandafter{\romannumeral3} An Example of Recomposition for 3-qubit case}
Consider that the correlation tensor $\mathcal R$ has the elements
$t_{033}=t_{303}=t_{330}=1$ (where $t_{000}=1$ is ignored). Noting that $t_{330}$, $t_{303}$ and $t_{033}$ are positive, and checking Tab.\uppercase\expandafter{\romannumeral2} in the main text, we have
\begin{small}
\begin{equation}
  \begin{aligned}
     &t_{330}\sigma_3\otimes\sigma_3\otimes\mathcal I + t_{303}\sigma_3\otimes\mathcal I\otimes\sigma_3 + t_{033}\mathcal I\otimes\sigma_3\otimes\sigma_3 \\
    =&\frac{t_{330}}{4}[(000)+(001)+(110)+(111)] +\frac{t_{303}}{4}[(000)+(010)+(101)+(111)] \\
     &+\frac{t_{033}}{4}[(000)+(100)+(011)+(111)] \\
    =&\frac{3}{4}[(000)+(111)]+ \frac{1}{4}[(001)+(010)+(011)+(100)+(101)+(110)] \\
    =&\frac{1}{2}(000)+\frac{1}{2}(111) :=l_{000}(000)+l_{111}(111),
  \end{aligned}
\end{equation}
\end{small}
where $l_{j_1 j_2 j_3}$ are positive, the sum of $l_{j_1 j_2 j_3}$ is the strength of the corresponding $t_{i_1i_2i_3}$. The above result shows that $t_{333}$ has a hidden strength with $t_{333}=|\frac{1}{2}\times1|+|\frac{1}{2}\times (-1)|=1$ from the columns $(000)$, $(111)$ and the least row of Tab.\uppercase\expandafter{\romannumeral1} in the main text. Because of the completeness of the SU-group, $\sum_{j_1,j_2,j_3=1}^{3} (j_1 j_2 j_3) = 0$. If $\sum l_{j_1 j_2 j_3} \leq 1$, $\frac{1}{8} [\mathcal I + \sum l_{j_1 j_2 j_3} (j_1 j_2 j_3)]$ is a feasible composition, and $\sum l_{j_1 j_2 j_3} = \sum_m |p_m t^{(m)}_{i_i i_2 i_3}|$ obviously.

To ensure that the recomposition does not interference the result of entanglement criterion, we need to begin the recomposition with the intersection of non-global elements.
For example, the $\sigma_3\otimes\sigma_3\otimes\mathcal I$ could be composed by $\sigma_3\otimes\sigma_3\otimes\sigma_1$, $\sigma_3\otimes\sigma_3\otimes\sigma_2$ and $\sigma_3\otimes\sigma_3\otimes\sigma_3$. The $\sigma_3\otimes\mathcal I\otimes\sigma_3$ and $\mathcal I\otimes\sigma_3\otimes\sigma_3$ are similar as $\sigma_3\otimes\sigma_3\otimes\mathcal I$. There will be 3 sets and each set has 3 elements. If we need to recompose $\sigma_3\otimes\sigma_3\otimes\mathcal I$, $\sigma_3\otimes\mathcal I\otimes\sigma_3$ and $\mathcal I\otimes\sigma_3\otimes\sigma_3$, the intersection set element $\sigma_3\otimes\sigma_3\otimes\sigma_3$ has the first priority. Considering the local hidden variable theory, the used non-global parts, which are $\sigma_3\otimes\sigma_3\otimes\mathcal I$, $\sigma_3\otimes\mathcal I\otimes\sigma_3$ and $\mathcal I\otimes\sigma_3\otimes\sigma_3$ should not be used twice in later processing. Note that if $t_{333}\neq0$, it should participate in the decomposition processing.

\section{\uppercase\expandafter{\romannumeral4} An Example of Rearrangement for 3-qubit Tensor $\mathcal S$}
For 3-qubit case, with the fixed $i_1$ and $i_2$, we have three $3\times 3$ matrices. Suppose these sets of singular values have the forms, $\{s^1_{i_3=1},s^2_{i_3=1},s^3_{i_3=1}\}_{I_1I_21}$, $\{s^1_{i_3=2},s^2_{i_3=2},s^3_{i_3=2}\}_{I_1I_22}$ and $\{s^1_{i_3=3},s^2_{i_3=3},s^3_{i_3=3}\}_{I_1I_23}$, respectively. The elements of the tensor $\mathcal S$ will be
\begin{equation}
\begin{aligned}
  \mathcal{S}_{::1}&=\left(
                      \begin{array}{ccc}
                        s^1_{i_3=1} &  0 & 0 \\
                        0 & s^2_{i_3=1} & 0 \\
                        0 &  0 & s^3_{i_3=1} \\
                      \end{array}
                    \right), \\
  \mathcal{S}_{::2}&=\left(
                      \begin{array}{ccc}
                        0 &s^1_{i_3=2} & 0 \\
                        0 & 0 & s^2_{i_3=2} \\
                        s^3_{i_3=2} & 0 & 0 \\
                      \end{array}
                    \right), \\
  \mathcal{S}_{::3}&=\left(
                      \begin{array}{ccc}
                        0 & 0 & s^1_{i_3=3} \\
                        s^2_{i_3=3} & 0 & 0 \\
                        0 & s^3_{i_3=3} & 0 \\
                      \end{array}
                    \right),
  \end{aligned}
\end{equation}
where $\mathcal{S}_{::i_3}$ represents the tensor slice with $i_3$. The rearrangement is non-unique, but it dose not influence our conclusion.

\section{\uppercase\expandafter{\romannumeral5} Examples of Multi-qubit GHZ State}
The 3-qubit GHZ state has the form, $|\psi\rangle=\frac{1}{\sqrt 2}(|000\rangle+|111\rangle)$, whose the tensor $\mathcal R$ is given by
\begin{equation}
  \begin{aligned}
  \mathcal R_{::0}=\left(
                     \begin{array}{cccc}
                       1 & 0 & 0 & 0 \\
                       0 & 0 & 0 & 0 \\
                       0 & 0 & 0 & 0 \\
                       0 & 0 & 0 & 1 \\
                     \end{array}
                   \right), \,\,\,
  \mathcal R_{::1}=\left(
                     \begin{array}{cccc}
                       0 & 0 &  0 & 0 \\
                       0 & 1 &  0 & 0 \\
                       0 & 0 & -1 & 0 \\
                       0 & 0 &  0 & 0 \\
                     \end{array}
                   \right), \\
  \mathcal R_{::2}=\left(
                     \begin{array}{cccc}
                       0 &  0 &  0 & 0 \\
                       0 &  0 & -1 & 0 \\
                       0 & -1 &  0 & 0 \\
                       0 &  0 &  0 & 0 \\
                     \end{array}
                   \right), \,\,\,
  \mathcal R_{::3}=\left(
                     \begin{array}{cccc}
                       0 & 0 & 0 & 1 \\
                       0 & 0 & 0 & 0 \\
                       0 & 0 & 0 & 0 \\
                       1 & 0 & 0 & 0 \\
                     \end{array}
                   \right),
  \end{aligned}
\end{equation}
where $\mathcal R_{::k}$ means the slice of $i_3=k$. $t_{033}=t_{303}=t_{330}=1$ and other $t_{i_133},t_{3i_23},t_{33i_3}$ are $0$. Thus we take $t_{333}$ as a decomposing base. From Tab.\uppercase\expandafter{\romannumeral2} in the main text, we have
\begin{equation}
  \begin{aligned}
    &t_{330}\sigma_3\otimes\sigma_3\otimes\mathcal I + t_{303}\sigma_3\otimes\mathcal I\otimes\sigma_3 + t_{033}\mathcal I\otimes\sigma_3\otimes\sigma_3 \\
      =&\frac{1}{2}(000)+\frac{1}{2}(111) =l_{000}(000)+l_{111}(111).
  \end{aligned}
\end{equation}
Since $l_{000}+l_{111} \leq 1$, $t_{\text{add}.333}=1$ and
\begin{equation}
  \mathcal{T}'_{::1}=\left(
                       \begin{array}{ccc}
                         1 &  0 & 0 \\
                         0 & -1 & 0 \\
                         0 &  0 & 0 \\
                       \end{array}
                     \right),
  \mathcal{T}'_{::2}=\left(
                       \begin{array}{ccc}
                         0 &-1 & 0 \\
                        -1 & 0 & 0 \\
                         0 & 0 & 0 \\
                       \end{array}
                     \right),
  \mathcal{T}'_{::3}=\left(
                       \begin{array}{ccc}
                         0 & 0 & 0 \\
                         0 & 0 & 0 \\
                         0 & 0 & 0 \\
                       \end{array}
                     \right).
\end{equation}
It is obvious that $t'_{k_1i_2i_3}=t'_{i_1k_2i_3}=t'_{i_1i_2k_3}=0$ if $t'_{i_1i_2i_3} \neq 0$, where $k=1,2,3$ and $k_n\neq i_n$. Thus $s_{i_1i_2i_3}=|t'_{i_1i_2i_3}|$.
$\mathcal T_{\text{add}}$ only has one element $t_{\text{add}.333}=1$. Hence, $S=5$.

Consider the state with a white noise of strength $q$, $\hat{\rho}=(1-q)\rho+q\frac{\mathcal I}{8}$. The tensor $\mathcal T'$ becomes $\hat{\mathcal T}'=(1-q) \mathcal T'$. In this case, $\hat{\rho}$ is an entangled state when $q<\frac{4}{5}$, a result that is the same as the one from PPT criterion.

Moreover, we can give a separable state decomposition when $q=\frac{4}{5}$, $p_1=p_2=\cdots=p_5=\frac{1}{5}$ with
\begin{equation}
  \begin{aligned}
    \rho_1&= \frac{1}{8}(\mathcal I +\sigma_1\otimes\sigma_1\otimes\sigma_1),~~~
    \rho_2 = \frac{1}{8}(\mathcal I -\sigma_2\otimes\sigma_2\otimes\sigma_1), \\
    \rho_3&= \frac{1}{8}(\mathcal I -\sigma_2\otimes\sigma_1\otimes\sigma_2),~~~
    \rho_4 = \frac{1}{8}(\mathcal I -\sigma_1\otimes\sigma_2\otimes\sigma_2), \\
    \rho_5&= \frac{1}{8}(\mathcal I +\sigma_3\otimes\sigma_3\otimes\mathcal I +\sigma_3\otimes\mathcal I\otimes\sigma_3 +\mathcal I \otimes\sigma_3\otimes\sigma_3).
  \end{aligned}
\end{equation}
Set $|\pm\rangle=\frac{1}{\sqrt{2}}(|0\rangle\pm|1\rangle$ and $|\tilde{\pm}\rangle=\frac{1}{\sqrt{2}}(|0\rangle\pm \text{i}|1\rangle)$. We can write the above ensemble in a pure state form with $p_1=p_2=\cdots=p_{16}=\frac{1}{20}$, $p_{17}=p_{18}=\frac{1}{10}$ and
\begin{equation}
  \begin{aligned}
    &|\psi_{1}\rangle=|\!+\!++\rangle, \quad |\psi_{2}\rangle=|\!+\!--\rangle, \quad |\psi_{3}\rangle=|\!-\!+-\rangle, \quad |\psi_{4}\rangle=|\!-\!-+\rangle, \\
    &|\psi_{5}\rangle=|\tilde{-}\tilde{+}+\rangle, \quad |\psi_{6}\rangle=|\tilde{+}\tilde{-}+\rangle, \quad |\psi_{7}\rangle=|\tilde{+}\tilde{+}-\rangle, \quad |\psi_{8}\rangle=|\tilde{-}\tilde{-}-\rangle, \\
    &|\psi_{9}\rangle=|\tilde{-}\!+\!\tilde{+}\rangle, \quad |\psi_{10}\rangle=|\tilde{+}\!-\!\tilde{+}\rangle, \quad |\psi_{11}\rangle=|\tilde{+}\!+\!\tilde{-}\rangle, \quad |\psi_{12}\rangle=|\tilde{-}\!-\!\tilde{-}\rangle, \\
    &|\psi_{13}\rangle=|\!-\!\tilde{+}\tilde{+}\rangle, \quad |\psi_{14}\rangle=|\!+\!\tilde{-}\tilde{+}\rangle, \quad |\psi_{15}\rangle=|\!+\!\tilde{+}\tilde{-}\rangle, \quad |\psi_{16}\rangle=|\!-\!\tilde{-}\tilde{-}\rangle, \\
    &|\psi_{17}\rangle=|000\rangle, \quad |\psi_{18}\rangle=|111\rangle.
  \end{aligned}
\end{equation}

Furthermore, we get the value of $S$ for the $N$-qubit GHZ state,
\begin{equation}
  S_N=2^{N-1}+1.
\end{equation}
Any state of the form $\rho=(1-q)|GHZ\rangle\langle GHZ|+q\frac{\mathcal I}{2^N}$ is entangled when $q<\frac{2^{N-1}}{2^{N-1}+1}$. This result proves the necessity and sufficiency of many previous entanglement criteria for GHZ state.

\section{\uppercase\expandafter{\romannumeral6} The case of the mixed $3$-qubit GHZ state}
It is well-known that the $3$-qubit system could be spanned by $2^3$ pure states as the base vectors.
There are 8 GHZ-type states,
  \begin{align}
    |\psi_1\rangle &= \frac{|000\rangle+|111\rangle}{\sqrt 2}, \,\,\, |\psi_2\rangle = \frac{|000\rangle-|111\rangle}{\sqrt 2}, \nonumber \\
    |\psi_3\rangle &= \frac{|001\rangle+|110\rangle}{\sqrt 2}, \,\,\, |\psi_4\rangle = \frac{|001\rangle-|110\rangle}{\sqrt 2}, \nonumber \\
    |\psi_5\rangle &= \frac{|010\rangle+|101\rangle}{\sqrt 2}, \,\,\, |\psi_6\rangle = \frac{|010\rangle-|101\rangle}{\sqrt 2}, \nonumber \\
    |\psi_7\rangle &= \frac{|011\rangle+|100\rangle}{\sqrt 2}, \,\,\, |\psi_8\rangle = \frac{|011\rangle-|100\rangle}{\sqrt 2}.
  \end{align}
Consider the state $\rho=\sum_k p_k |\psi_k\rangle\langle \psi_k|$. For this state,
the tensor $\mathcal{T}$ only has 4 elements,
\begin{equation}
  \begin{aligned}
    t_{111} &= p_1 -p_2 +p_3 -p_4 +p_5 -p_6 +p_7 -p_8, \\
    t_{122} &=-p_1 +p_2 +p_3 -p_4 +p_5 -p_6 -p_7 +p_8, \\
    t_{212} &=-p_1 +p_2 +p_3 -p_4 -p_5 +p_6 +p_7 -p_8, \\
    t_{221} &=-p_1 +p_2 -p_3 +p_4 +p_5 -p_6 +p_7 -p_8.
  \end{aligned}
\end{equation}
For $t_{033}$, $t_{303}$ and $t_{330}$, we can use $t_{333}$ as a base.
From Tab.\uppercase\expandafter{\romannumeral2} in the main text, we have
\begin{equation}
  \begin{aligned}
    &\frac{1}{4}(p_1+p_2+p_3+p_4)[(000)+(001)+(110)+(111)] \\
    &+\frac{1}{4}(p_5+p_6+p_7+p_8)[(010)+(011)+(100)+(101)] \\
    &+\frac{1}{4}(p_1+p_2+p_5+p_6)[(000)+(010)+(101)+(111)] \\
    &+\frac{1}{4}(p_3+p_4+p_7+p_8)[(001)+(011)+(100)+(110)] \\
    &+\frac{1}{4}(p_1+p_2+p_7+p_8)[(000)+(100)+(011)+(111)] \\
    &+\frac{1}{4}(p_3+p_4+p_5+p_6)[(001)+(101)+(010)+(110)] \\
    =&(p_1+p_2)[\frac{1}{2}(000)+\frac{1}{2}(111)] + (p_3+p_4)[\frac{1}{2}(001)+\frac{1}{2}(110)] \\
     &+ (p_5+p_6)[\frac{1}{2}(010)+\frac{1}{2}(101)] + (p_7+p_8)[\frac{1}{2}(011)+\frac{1}{2}(100)].
  \end{aligned}
\end{equation}
It means that
\begin{equation}
  t_{\text{add}.333}=1-4\min\{p_1+p_2,\, p_3+p_4,\, p_5+p_6,\, p_7+p_8\}\geq 0,
\end{equation}
and $S$ is given by
\begin{equation}
  S=|t_{111}|+|t_{122}|+|t_{212}|+|t_{221}|+t_{\text{add}.333},
\end{equation}
from which we can easily judge whether $\rho$ is entangled or not according to the probability distribution for orthogonal GHZ-type states.

\section{\uppercase\expandafter{\romannumeral7} Calculation for W States}
The 3-qubit W state has the form, $|\psi\rangle=\frac{1}{\sqrt{3}}(|001\rangle+|010\rangle+|100\rangle)$. The related tensor $\mathcal R$ is given by
\begin{equation}
  \begin{aligned}
  \mathcal R_{::0}=\left(
                     \begin{array}{cccc}
                       1 & 0 & 0 & \frac{1}{3} \\
                       0 & \frac{2}{3} & 0 & 0 \\
                       0 & 0 & \frac{2}{3} & 0 \\
                       \frac{1}{3} & 0 & 0 & -\frac{1}{3} \\
                     \end{array}
                   \right), \,\,\,
  \mathcal R_{::1}=\left(
                     \begin{array}{cccc}
                       0 & \frac{2}{3} & 0 & 0 \\
                       \frac{2}{3} & 0 & 0 & \frac{2}{3} \\
                       0 & 0 & 0 & 0 \\
                       0 & \frac{2}{3} & 0 & 0 \\
                     \end{array}
                   \right), \\
  \mathcal R_{::2}=\left(
                     \begin{array}{cccc}
                       0 & 0 & \frac{2}{3} & 0 \\
                       0 & 0 & 0 & 0 \\
                       \frac{2}{3} & 0 & 0 & \frac{2}{3} \\
                       0 & 0 & \frac{2}{3} & 0 \\
                     \end{array}
                   \right), \,\,\,
  \mathcal R_{::3}=\left(
                     \begin{array}{cccc}
                       \frac{1}{3} & 0 & 0 & -\frac{1}{3} \\
                       0 & \frac{2}{3} & 0 & 0 \\
                       0 & 0 & \frac{2}{3} & 0 \\
                       -\frac{1}{3} & 0 & 0 & -1 \\
                     \end{array}
                   \right).
  \end{aligned}
\end{equation}
In this case, $t_{011}=t_{101}=t_{110}=\frac{2}{3}$, $t_{022}=t_{202}=t_{220}=\frac{2}{3}$, $t_{003}=t_{030}=t_{300}=\frac{1}{3}$ and $t_{033}=t_{303}=t_{330}=-\frac{1}{3}$. We find that $t_{003}$, $t_{030}$, $t_{300}$, $t_{033}$, $t_{303}$, $t_{330}$ and $t_{333}$ can be composed and included into $t_{333}=-1$ with $l_{001}=l_{010}=l_{100}=\frac{1}{3}$. For operating Tab.\uppercase\expandafter{\romannumeral1} and Tab.\uppercase\expandafter{\romannumeral2} in the main text, we can calculate and get that $t_{011}$, $t_{101}$, $t_{110}$ according to $t_{111}=\frac{2}{3}$ with $l_{+++}=l_{---}=\frac{1}{3}$. Similarly, $t_{022}$, $t_{202}$, $t_{220}$ according to $t_{222}=\frac{2}{3}$ with $l_{\tilde + \tilde + \tilde+}=l_{\tilde - \tilde - \tilde -}=\frac{1}{3}$ are also obtained. The subscripts of $l$ represent the quantum states $|\pm\rangle=\frac{|0\rangle\pm|1\rangle}{\sqrt{2}}$ and $|\tilde\pm\rangle=\frac{|0\rangle\pm \text{i}|1\rangle}{\sqrt{2}}$.

Note that $t_{030}$, $t_{101}$ and $t_{131}$ is also a feasible composition. But because of the local hidden variable theory, it will interfere another distribution of elements. Therefore, we usually begin this composition with the element of intersection which is generated by the sets number as many as possible.

Now we have the tensor $\mathcal T'$,
\begin{equation}
  \mathcal{T}'_{::1}=\left(
                       \begin{array}{ccc}
                         0 & 0 & \frac{2}{3} \\
                         0 & 0 & 0 \\
                         \frac{2}{3} &  0 & 0 \\
                       \end{array}
                     \right),~~~
  \mathcal{T}'_{::2}=\left(
                       \begin{array}{ccc}
                         0 & 0 & 0 \\
                         0 & 0 & \frac{2}{3} \\
                         0 & \frac{2}{3} & 0 \\
                       \end{array}
                     \right),~~~
  \mathcal{T}'_{::3}=\left(
                       \begin{array}{ccc}
                         \frac{2}{3} & 0 & 0 \\
                         0 & \frac{2}{3} & 0 \\
                         0 & 0 & -1 \\
                       \end{array}
                     \right),
\end{equation}
and the tensor $\mathcal T'_{\text{add}}$,
\begin{equation}
    \mathcal{T}'_{\text{add}.::1}=\left(
                                     \begin{array}{ccc}
                                       \frac{2}{3} & 0 & 0 \\
                                       0 & 0 & 0 \\
                                       0 & 0 & 0 \\
                                     \end{array}
                                   \right),~~~
    \mathcal{T}'_{\text{add}.::2}=\left(
                                    \begin{array}{ccc}
                                      0 & 0 & 0 \\
                                      0 & \frac{2}{3} & 0 \\
                                      0 & 0 & 0 \\
                                    \end{array}
                                  \right),~~~
    \mathcal{T}'_{\text{add}.::3}=\left(
                                     \begin{array}{ccc}
                                       0 & 0 & 0 \\
                                       0 & 0 & 0 \\
                                       0 & 0 & 0 \\
                                     \end{array}
                                   \right),
\end{equation}
and $S=\frac{19}{3}$. The separable state ensemble is shown in the main text.

The 4-qubit W state $|\psi\rangle=\frac{1}{2}(|0001\rangle+|0010\rangle+|0100\rangle+|1000\rangle)$ has the tensor $\mathcal R$ with tensor elements
\begin{align}
  &t_{0003}=\frac{1}{2},\quad t_{0030}=\frac{1}{2},\quad t_{0300}=\frac{1}{2},\quad t_{3000}=\frac{1}{2}, \nonumber \\
  &t_{0333}=-\frac{1}{2},\quad t_{3033}=-\frac{1}{2},\quad t_{3303}=-\frac{1}{2},\quad t_{3330}=-\frac{1}{2},\quad t_{3333}=-1, \nonumber \\
  &t_{0011}=\frac{1}{2},\quad t_{0101}=\frac{1}{2},\quad t_{0110}=\frac{1}{2},\quad t_{1001}=\frac{1}{2}, \nonumber \\
  &t_{1010}=\frac{1}{2},\quad t_{1100}=\frac{1}{2},\quad t_{0022}=\frac{1}{2},\quad t_{0202}=\frac{1}{2}, \nonumber \\
  &t_{0220}=\frac{1}{2},\quad t_{2002}=\frac{1}{2},\quad t_{2020}=\frac{1}{2},\quad t_{2200}=\frac{1}{2}, \nonumber \\
  &t_{1133}=\frac{1}{2},\quad t_{1103}=\frac{1}{2},\quad t_{1130}=\frac{1}{2}, \nonumber \\
  &t_{1313}=\frac{1}{2},\quad t_{1013}=\frac{1}{2},\quad t_{1310}=\frac{1}{2}, \nonumber \\
  &t_{1331}=\frac{1}{2},\quad t_{1031}=\frac{1}{2},\quad t_{1301}=\frac{1}{2}, \nonumber \\
  &t_{3113}=\frac{1}{2},\quad t_{0113}=\frac{1}{2},\quad t_{3110}=\frac{1}{2}, \nonumber \\
  &t_{3131}=\frac{1}{2},\quad t_{0131}=\frac{1}{2},\quad t_{3101}=\frac{1}{2}, \nonumber \\
  &t_{3311}=\frac{1}{2},\quad t_{0311}=\frac{1}{2},\quad t_{3011}=\frac{1}{2}, \nonumber \\
  &t_{2233}=\frac{1}{2},\quad t_{2203}=\frac{1}{2},\quad t_{2230}=\frac{1}{2}, \nonumber \\
  &t_{2323}=\frac{1}{2},\quad t_{2023}=\frac{1}{2},\quad t_{2320}=\frac{1}{2}, \nonumber \\
  &t_{2332}=\frac{1}{2},\quad t_{2032}=\frac{1}{2},\quad t_{2302}=\frac{1}{2}, \nonumber \\
  &t_{3223}=\frac{1}{2},\quad t_{0223}=\frac{1}{2},\quad t_{3220}=\frac{1}{2}, \nonumber \\
  &t_{3232}=\frac{1}{2},\quad t_{0232}=\frac{1}{2},\quad t_{3202}=\frac{1}{2}, \nonumber \\
  &t_{3322}=\frac{1}{2},\quad t_{0322}=\frac{1}{2},\quad t_{3022}=\frac{1}{2}.
\end{align}
When $q=\frac{20}{21}$, we can also provide a separable state ensemble, which is given by
$p_1=p_2=p_3=\frac{1}{21}$, $p_4=p_5=\cdots=p_{39}=\frac{1}{42}$ and
  \begin{align}
    \rho_{1}=&\frac{1}{16}(\mathcal I_{16} +\frac{1}{2}\mathcal{III}\sigma_3 +\frac{1}{2}\mathcal{II}\sigma_3\mathcal I+\frac{1}{2}\mathcal I\sigma_3\mathcal{II} + \frac{1}{2}\sigma_3\mathcal{III} \nonumber \\
           &-\frac{1}{2}\mathcal I \sigma_3\sigma_3\sigma_3 -\frac{1}{2}\sigma_3\mathcal I \sigma_3\sigma_3 -\frac{1}{2}\sigma_3\sigma_3\mathcal I\sigma_3 -\frac{1}{2}\sigma_3\sigma_3\sigma_3\mathcal I -\sigma_3\sigma_3\sigma_3\sigma_3), \nonumber \\
    \rho_{2}=&\frac{1}{16}(\mathcal I_{16} +\frac{1}{2}\mathcal{II}\sigma_1\sigma_1 +\frac{1}{2}\mathcal I\sigma_1\mathcal I\sigma_1 +\frac{1}{2}\mathcal I\sigma_1\sigma_1\mathcal I +\frac{1}{2}\sigma_1\mathcal{II}\sigma_1 +\frac{1}{2}\sigma_1\mathcal I\sigma_1\mathcal I +\frac{1}{2}\sigma_1\sigma_1\mathcal{II}), \nonumber \\
    \rho_{3}=&\frac{1}{16}(\mathcal I_{16} +\frac{1}{2}\mathcal{II}\sigma_2\sigma_2 +\frac{1}{2}\mathcal I\sigma_2\mathcal I\sigma_2 +\frac{1}{2}\mathcal I\sigma_2\sigma_2\mathcal I +\frac{1}{2}\sigma_2\mathcal{II}\sigma_2 +\frac{1}{2}\sigma_2\mathcal I\sigma_2\mathcal I +\frac{1}{2}\sigma_2\sigma_2\mathcal{II}), \nonumber \\
    \rho_{4}=&\frac{1}{16}(\mathcal I_{16} +\frac{1}{2}\sigma_1\sigma_1\sigma_3\sigma_3),
    \rho_{5}=\frac{1}{16}(\mathcal I_{16} +\frac{1}{2}\sigma_1\sigma_1\mathcal I\sigma_3),
    \rho_{6}=\frac{1}{16}(\mathcal I_{16} +\frac{1}{2}\sigma_1\sigma_1\sigma_3\mathcal I), \nonumber \\
    \rho_{7}=&\frac{1}{16}(\mathcal I_{16} +\frac{1}{2}\sigma_1\sigma_3\sigma_1\sigma_3),
    \rho_{8}=\frac{1}{16}(\mathcal I_{16} +\frac{1}{2}\sigma_1\mathcal I\sigma_1\sigma_3),
    \rho_{9}=\frac{1}{16}(\mathcal I_{16} +\frac{1}{2}\sigma_1\sigma_3\sigma_1\mathcal I), \nonumber \\
    \rho_{10}=&\frac{1}{16}(\mathcal I_{16} +\frac{1}{2}\sigma_1\sigma_3\sigma_3\sigma_1),
    \rho_{11}=\frac{1}{16}(\mathcal I_{16} +\frac{1}{2}\sigma_1\mathcal I\sigma_3\sigma_1),
    \rho_{12}=\frac{1}{16}(\mathcal I_{16} +\frac{1}{2}\sigma_1\sigma_3\mathcal I\sigma_1), \nonumber \\
    \rho_{13}=&\frac{1}{16}(\mathcal I_{16} +\frac{1}{2}\sigma_3\sigma_1\sigma_1\sigma_3),
    \rho_{14}=\frac{1}{16}(\mathcal I_{16} +\frac{1}{2}\mathcal I\sigma_1\sigma_1\sigma_3),
    \rho_{15}=\frac{1}{16}(\mathcal I_{16} +\frac{1}{2}\sigma_3\sigma_1\sigma_1\mathcal I), \nonumber \\
    \rho_{16}=&\frac{1}{16}(\mathcal I_{16} +\frac{1}{2}\sigma_3\sigma_1\sigma_3\sigma_1),
    \rho_{17}=\frac{1}{16}(\mathcal I_{16} +\frac{1}{2}\mathcal I\sigma_1\sigma_3\sigma_1),
    \rho_{18}=\frac{1}{16}(\mathcal I_{16} +\frac{1}{2}\sigma_3\sigma_1\mathcal I\sigma_1), \nonumber \\
    \rho_{19}=&\frac{1}{16}(\mathcal I_{16} +\frac{1}{2}\sigma_3\sigma_3\sigma_1\sigma_1),
    \rho_{20}=\frac{1}{16}(\mathcal I_{16} +\frac{1}{2}\mathcal I\sigma_3\sigma_1\sigma_1),
    \rho_{21}=\frac{1}{16}(\mathcal I_{16} +\frac{1}{2}\sigma_3\mathcal I\sigma_1\sigma_1), \nonumber \\
    \rho_{22}=&\frac{1}{16}(\mathcal I_{16} +\frac{1}{2}\sigma_2\sigma_2\sigma_3\sigma_3),
    \rho_{23}=\frac{1}{16}(\mathcal I_{16} +\frac{1}{2}\sigma_2\sigma_2\mathcal I\sigma_3),
    \rho_{24}=\frac{1}{16}(\mathcal I_{16} +\frac{1}{2}\sigma_2\sigma_2\sigma_3\mathcal I), \nonumber \\
    \rho_{25}=&\frac{1}{16}(\mathcal I_{16} +\frac{1}{2}\sigma_2\sigma_3\sigma_2\sigma_3),
    \rho_{26}=\frac{1}{16}(\mathcal I_{16} +\frac{1}{2}\sigma_2\mathcal I\sigma_2\sigma_3),
    \rho_{27}=\frac{1}{16}(\mathcal I_{16} +\frac{1}{2}\sigma_2\sigma_3\sigma_2\mathcal I), \nonumber \\
    \rho_{28}=&\frac{1}{16}(\mathcal I_{16} +\frac{1}{2}\sigma_2\sigma_3\sigma_3\sigma_2),
    \rho_{29}=\frac{1}{16}(\mathcal I_{16} +\frac{1}{2}\sigma_2\mathcal I\sigma_3\sigma_2),
    \rho_{30}=\frac{1}{16}(\mathcal I_{16} +\frac{1}{2}\sigma_2\sigma_3\mathcal I\sigma_2), \nonumber \\
    \rho_{31}=&\frac{1}{16}(\mathcal I_{16} +\frac{1}{2}\sigma_3\sigma_2\sigma_2\sigma_3),
    \rho_{32}=\frac{1}{16}(\mathcal I_{16} +\frac{1}{2}\mathcal I\sigma_2\sigma_2\sigma_3),
    \rho_{33}=\frac{1}{16}(\mathcal I_{16} +\frac{1}{2}\sigma_3\sigma_2\sigma_2\mathcal I), \nonumber \\
    \rho_{34}=&\frac{1}{16}(\mathcal I_{16} +\frac{1}{2}\sigma_3\sigma_2\sigma_3\sigma_2),
    \rho_{35}=\frac{1}{16}(\mathcal I_{16} +\frac{1}{2}\mathcal I\sigma_2\sigma_3\sigma_2),
    \rho_{36}=\frac{1}{16}(\mathcal I_{16} +\frac{1}{2}\sigma_3\sigma_2\mathcal I\sigma_2), \nonumber \\
    \rho_{37}=&\frac{1}{16}(\mathcal I_{16} +\frac{1}{2}\sigma_3\sigma_3\sigma_2\sigma_2),
    \rho_{38}=\frac{1}{16}(\mathcal I_{16} +\frac{1}{2}\mathcal I\sigma_3\sigma_2\sigma_2),
    \rho_{39}=\frac{1}{16}(\mathcal I_{16} +\frac{1}{2}\sigma_3\mathcal I\sigma_2\sigma_2),
  \end{align}
where $\mathcal I_{16}$ is the $16\times 16$ identity matrix, the tensor sign $\otimes$ among $\sigma_{1,2,3}$ and $\mathcal I$ has been ignored.

\section{\uppercase\expandafter{\romannumeral8} The Entanglement Robustness and Sudden Death}
It has been shown that the entanglement robustness of multipartite GHZ state, $|\psi\rangle_N=\frac{1}{\sqrt{2}}(|0\rangle^{\otimes N}+|1\rangle^{\otimes N})$, under local depolarized noise will be changed with the changing of the numbers of subsystems. For weak noise strength, the more the subsystems, the less robustness. However, with the increasing of noise strength, the entanglement of the states with the least subsystems will vanish firstly \cite{PhysRevLett.100.080501}.

The depolarized channel with strength $q$ is given by the Kraus operators,
\begin{equation}
  \begin{aligned}
    \mathcal E_1=\sqrt{1-\frac{3q}{4}}\left(
                               \begin{array}{cc}
                                 1 & 0 \\
                                 0 & 1 \\
                               \end{array}
                             \right),& \,\,\,
    ~~~\mathcal E_2=\sqrt{\frac{q}{4}}\left(
                            \begin{array}{cc}
                              0 & 1 \\
                              1 & 0 \\
                            \end{array}
                          \right), \\
    \mathcal E_3=\sqrt{\frac{q}{4}}\left(
                            \begin{array}{cc}
                              0 & -\text{i} \\
                              \text{i} & 0 \\
                            \end{array}
                          \right),& \,\,\,
    ~~~\mathcal E_4=\sqrt{\frac{q}{4}}\left(
                            \begin{array}{cc}
                              1 & 0 \\
                              0 & -1 \\
                            \end{array}
                          \right).
  \end{aligned}
\end{equation}
Under the depolarized channel a qubit state $\rho$ is mapped to be
$\mathcal E(\rho)=\sum_k \mathcal E_k \rho \mathcal E_k^\dagger$.

This phenomenon of entanglement sudden death has not been well illustrated. We can provide an explanation by using our method.
For the $N$-qubit GHZ state, we have $S_N=2^{N-1}+1$.
With the noise channel acting on the each qubit, the tensor $\mathcal T$ becomes $(1-q)^N \mathcal T$.
Benefited from the high symmetry of GHZ state, we can define a simple entanglement monotone for the $N$-qubit GHZ state,
\begin{equation}
\begin{aligned}
  E_N=&\max(\frac{\hat S_N-1}{S_N-1},0) \\
     =&\max(\frac{S_N(1-q)^N-1}{S_N-1},0),
\end{aligned}
\end{equation}
where $N$ is the number of qubits and $\hat S$ represents the minimal sum of non-locality. The denominator ensures that the maximal value of this monotone is equal to $1$.

Multipartite entanglement is based on the correlation among the qubits. So $E_N$ is defined as the minimal sum of multipartite correlation strength under local operations and convex combinations. It is slightly different from $S$ which represents the minimal value of the sum of probability $p$ with all possible convex combinations of separable states except for the completely mixed state.

\begin{figure}[htbp]
\centering
\includegraphics[width=0.50\textwidth]{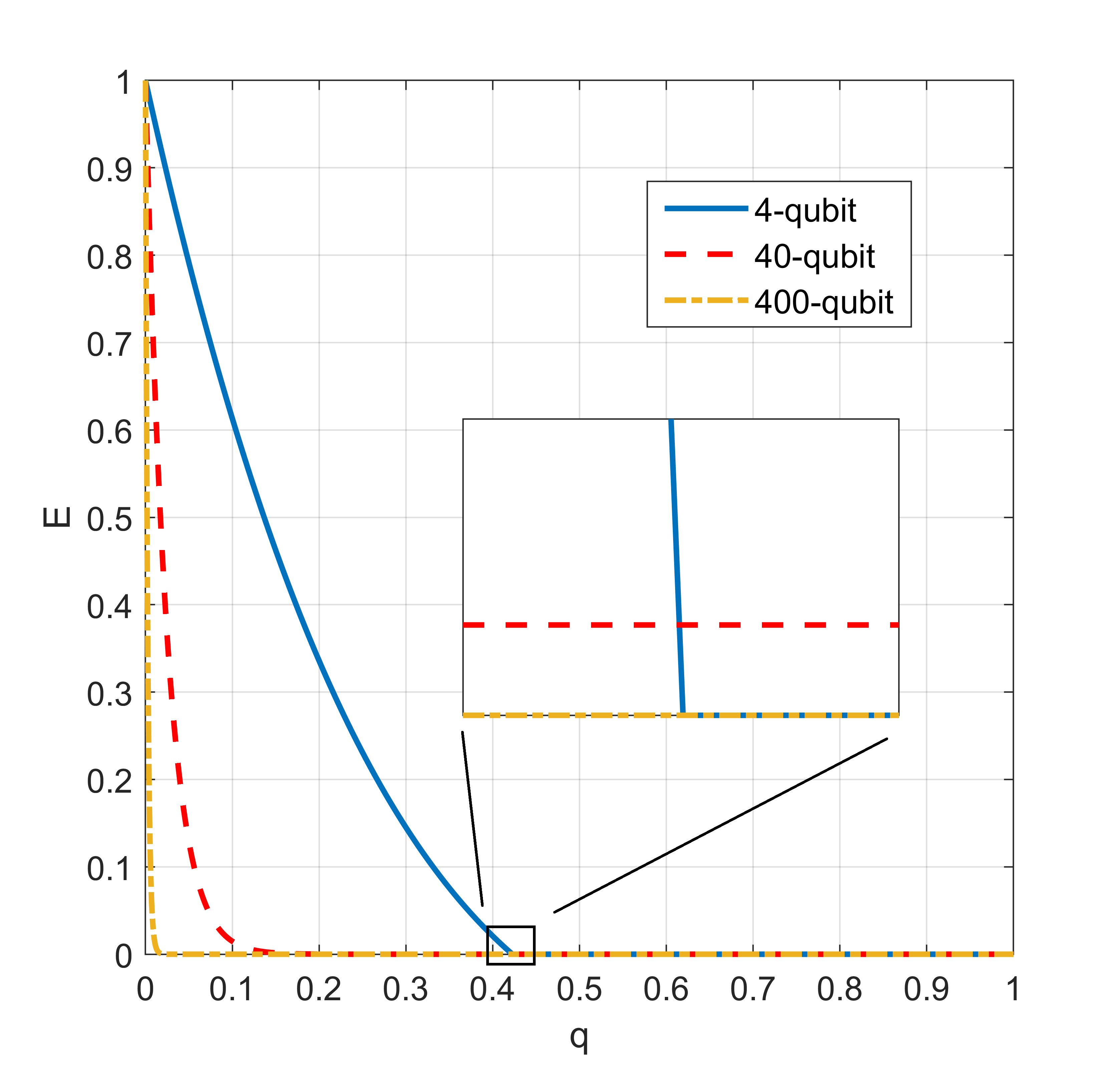}
\caption{The $E_N$ versus noise strength $q$ of local depolarized channel. The blue line, the red dash line and the yellow dot-dash line are for the $4$-qubit, the $40$-qubit and the $400$-qubit cases, respectively. The inset shows a magnification of the region in which $E_4$ vanishes. One has that $q|_{E_{400}=0}>q|_{E_{40}=0}>q|_{E_4=0}$.}
\label{Enof4-400}
\end{figure}

With $N$ as the power exponent of $1-q$, we have Fig.\ref{Enof4-400} with $N=4,40,400$. The behavior of the entanglement is the same as that in Ref.\cite{PhysRevLett.100.080501}. The power exponent $N$ is the reason why the entanglement robustness has such phenomenon.

\begin{figure}[htbp]
\centering
\includegraphics[width=0.95\textwidth]{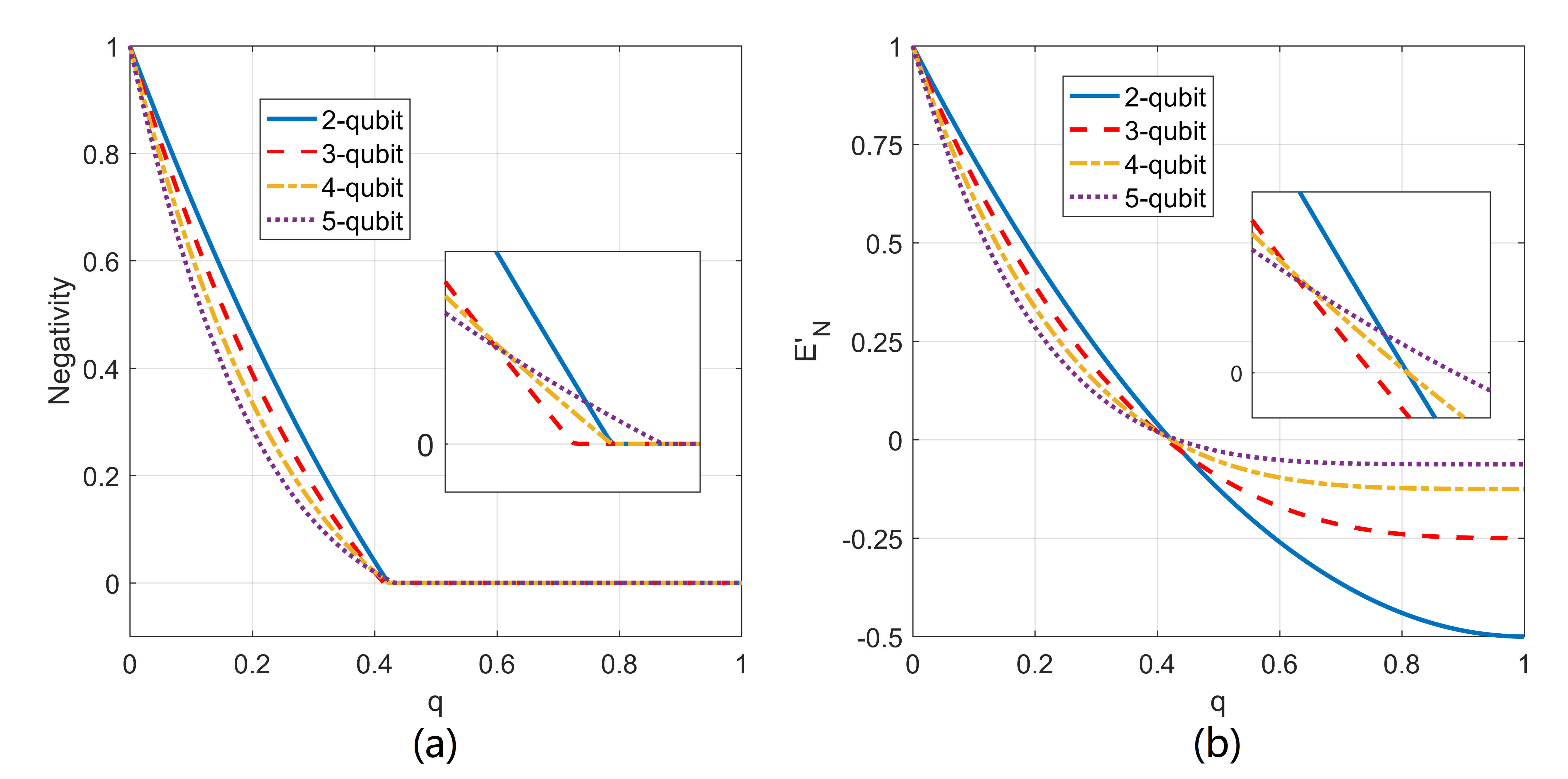}
\caption{The entanglement negativity (a) and $E'_N$ (b) versus noise strength $q$ with $N=2,3,4,5$. The blue, red dashed, yellow dot-dashed and purple dotted lines are for the $2$-qubit, $3$-qubit, $4$-qubit and $5$-qubit cases, respectively. The insets show the magnifications of the regions in which $q\in[0.39,0.44]$, negativity and $E'_N$ in the area of $[-0.01,0.04]$. In Figure (a) the negativity functions with respect to 2-qubit and 3-qubit have no crossing point. In Figure (b), the crossing point with respect to $2$-qubit and $3$-qubit is below $E'_N=0$.}
\label{Neg2345AndOurs}
\end{figure}

We show the negativity for the cases of $N=2,3,4,5$ in Fig.\ref{Neg2345AndOurs}(a). $N$-qubit system is divided as $k_1$ qubit(s) and $k_2$ qubit(s) $\{k_1,k_2\}$, where $k_1=\lfloor \frac{N}{2} \rfloor$ and $k_1+k_2=N$. The 2-qubit is divided as \{1,1\}, 3-qubit as $\{1,2\}$, 4-qubit as $\{2,2\}$ and 5-qubit as $\{2,3\}$, and the partial transpose acts on the first $k_1$ qubit(s). In fact, because of the high symmetry of GHZ state, the results do not depend on the ways of such dividing. Intuitively, the entanglement will terminates completely after a infinite interval of time, with a smoothly diminishing long-time tail. But there is a sudden change of slope where the negativity value is $0$. This phenomenon is called the sudden death of entanglement \cite{science.1142654}.
There are two questions with this result. Firstly, why the negativity functions of $N=2$ and $N=3$ have no crossing point. Secondly, is the ``sudden death'' real ``sudden''? These questions are answered better by our method.

In fact, the crossing point of $2$-qubit and $3$-qubit negativity functions is below $E=0$, see Fig.\ref{Neg2345AndOurs}(b). From $E'_N=\frac{S_N(1-q)^N-1}{S_N-1}$, one sees that the data above $E'=0$ is the same as the negativity for GHZ states in numerical results (5 significant digits), and the data below $E'=0$ represents that the state is separable.

\begin{figure}[htbp]
\centering
\includegraphics[width=0.50\textwidth]{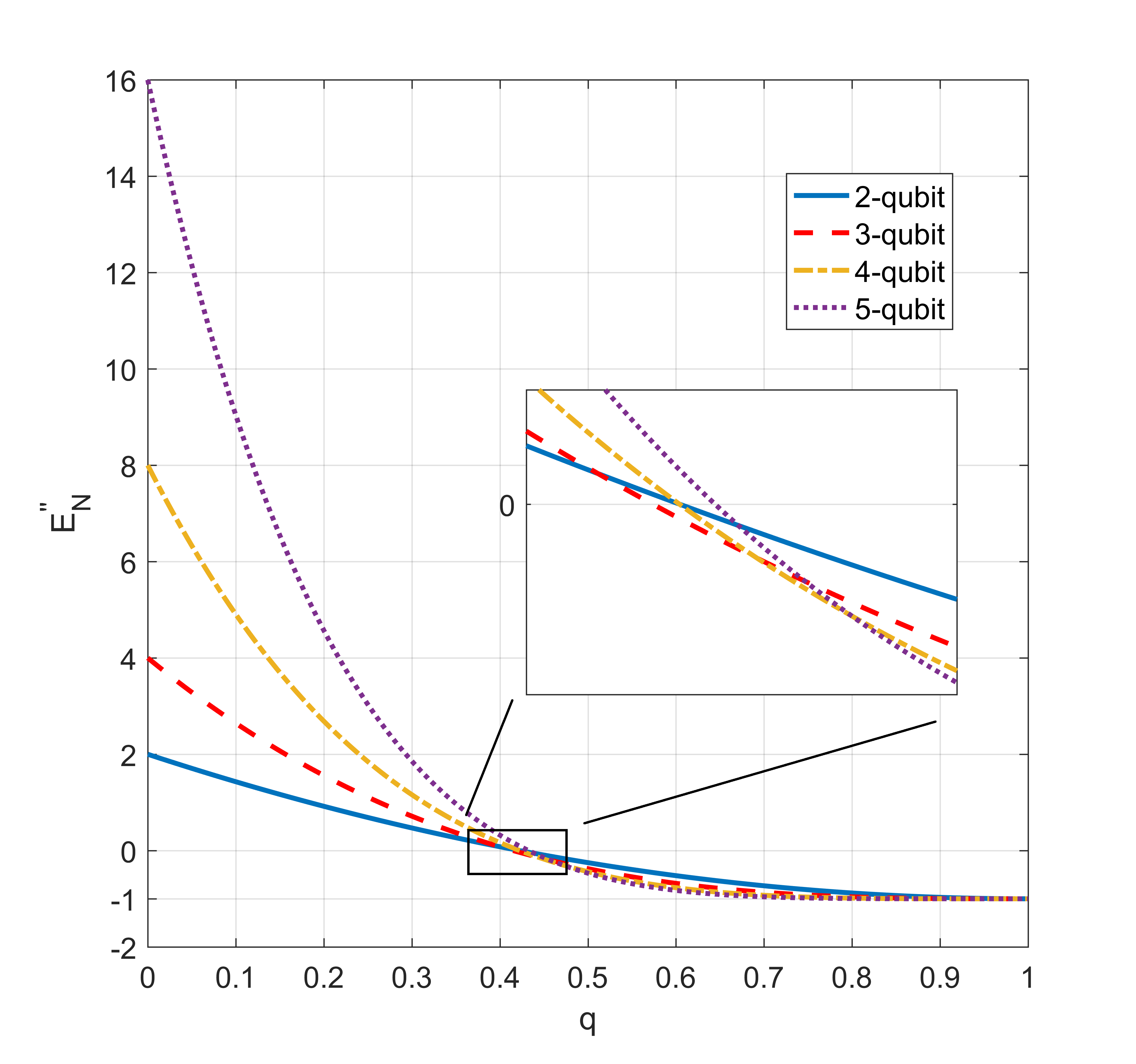}
\caption{The $E''_N$ versus noise strength $q$ with $N=2,3,4,5$. The inset shows a magnification of the region in which $q\in[0.38,0.5]$ and $E''_N \in [-0.5,0.3]$. One sees that the ``sudden death'' of entanglement is no longer ``sudden'' at all.}
\label{E''2345}
\end{figure}

If we remove the normalized constant of denominator and plot $E''_N=S_N(1-q)^N-1$ directly, the ``sudden death'' will no longer ``sudden'' at all (see Fig.\ref{E''2345}). The $E''_N$ is a continuous function between $q=0$ and $q=1$ with $E''_N|_{q=1}=-1$ and $\frac{\text d E''_N}{\text d q}|_{q=1}=0$. It is well known that there is a positive correlation between $1-q$ and $\exp(-\text i Ht)$, where $H$ is the Hamiltonian and $t$ is the time. It means that when $t\rightarrow \infty$, $E''_N+1\rightarrow 0$. Now $E''+1$ has a smooth exponential decay of $e$, which is 0 at infinity of $t$. However, the entanglement will vanish in the point of $E''_N=0$, which is the point of ``sudden death''. Hence, it is clear that the ``sudden death'' is due to that the negativity ignores the continuous change from entangled state to separable state.

The above results show that our method reals some hidden facts and explains the sudden death better. It reflects that our criterion is more accurate than the PPT criterion and has some advantages in physical manifestations.

\end{document}